\documentclass[aps,prl,twocolumn,showpacs,superscriptaddress,groupedaddress]{revtex4-1}
\usepackage[dvipdfmx]{graphicx}

\usepackage{stackrel,amssymb}
\usepackage{amsmath,chemarrow}

\usepackage[version=3]{mhchem}

\usepackage{color}

\usepackage{here}

\begin{document}

\title{Kinetic Selection of Template Polymer with Complex Sequences}
\author{Yoshiya J. Matsubara}
\email{yoshi@copmlex.c.u-tokyo.ac.jp, Tel.: +81-3-5454-6732}
\author{Kunihiko Kaneko}
\email{kaneko@complex.c.u-tokyo.ac.jp, Tel.: +81-3-5454-6746}
\affiliation{Graduate School of Arts and Sciences, The University of Tokyo, 3-8-1 Komaba, Meguro-ku, Tokyo 153-8902, Japan}
\date{\today}

\begin{abstract}
The emergence and maintenance of polymers with complex sequences pose a major question in the study of the origin of life. To answer this, we study a model polymerization reaction, where polymers are synthesized by stepwise ligation from two types of monomers, catalyzed by a long polymer as a template. Direct stochastic simulation and dynamical systems analysis reveal that the most dominant polymer sequence in a population successively changes depending on the flow rate of monomers to the system, with more complex sequences selected at a lower flow rate. We discuss the relevance of this kinetic sequence selection through nonequilibrium flow to the origin of complex polymers.
\end{abstract}

\maketitle

Schr\"{o}dinger, in his celebrated monograph ``What is Life'' \cite{schrodinger1943life}, recognized that the essence of a heredity-carrier lies in aperiodic crystals, as was soon confirmed by the elucidation of DNA structure. In general, biopolymers consist of different kinds of monomers, constituting an aperiodic complex sequence, as is important not only for genetic information, but also for the catalytic function. 

With regard to the origin of life or self-replication, catalytic polymers are required, as has also been investigated in recent experiments \cite{von1986self, lincoln2009self, krammer2012thermal, vaidya2012spontaneous, sadownik2016diversification}. In their study, polymers with complex sequences had to be synthesized and preserved, to encode a large amount of information and for catalytic functions \cite{carothers2004informational, Derr2012}.
Hence, it is important to uncover the condition that allows for the generation of a polymer with a complex monomer sequence. 

Polymers such as RNA are replicated using other polymers as templates. In the prebiotic world, complex monomer sequences therefore have been synthesized by reactions with templates as catalysts, which are also synthesized via such template reaction. Thus, the polymerization processes are autocatalytic in nature, as has been extensively investigated \cite{eigen1977principle, farmer1986autocatalytic, szabo2002silico, kaneko2005recursive, giri2012origin,  Matsubara2016, guseva2017foldamer, kinsler2017prebiotic}, including theoretical models that explicitly consider template-like self-replicating polymerization \cite{anderson1983suggested, Stein1984, fernando2007stochastic,  nowak2008prevolutionary, Ohtsuki2009, Obermayer2011,  Derr2012, Tanaka2014, Tkachenko2015} and some synthesis experiments \cite{sadownik2016diversification, he2016viscous, toyabe2018templated}.

If there exist multiple catalytic polymers with sufficient lengths, then the autocatalytic process for their synthesis can select one of such polymers.
Here we study the dependence of the selected sequence on the flow rate of monomers.
The replication of polymers needs a supply of monomers to compensate polymer degradation. Such a nonequilibrium open condition could be provided by a hydrothermal vent at the origin of life and by a chemostat condition in a laboratory experiment \cite{novick1950description}.
Despite the recognition that such nonequilibrium flow is essential to the emergence and maintenance of life, its influence on the selection of sequence has not been systematically investigated. 

The selection of a specific polymer by catalytic reactions studied so far \cite{anderson1983suggested, Stein1984, fernando2007stochastic,  nowak2008prevolutionary, Ohtsuki2009, Obermayer2011,  Derr2012, Tanaka2014, Tkachenko2015} does not show any dependence on external conditions. Ref \cite{kinsler2017prebiotic} reported the dependence of the selected polymer sequence on the bias in the monomer components. In contrast, here, we study a template-catalyst polymerization model by a complementary sequence, and demonstrate that the selected sequence of the template polymer depends critically on the flow rate, and more complex sequences are selected as the flow rate is decreased.
This selection mechanism is the {\sl kinetic}, rather than {\sl energetic}; selection of complex sequences (that include a variety of subsequences, which are defined later) without any specific design of energy dependence, as a result of changing the rate of external supply of monomers.
As will be analyzed by dynamical systems and combinatorial analysis, complex sequences that are synthesized via diverse kinetic pathways from monomers are selected for a low flow rate.  The generality of this selection mechanism will be discussed with possible relevance to prebiotic polymer synthesis.

We consider a synthetic reaction of polymers, which consist of two kinds of monomers, denoted as 0 and 1, where each polymerization is catalyzed by a long template polymer. Thus, all polymer species are described by the binary integer $s$ of length $l$ (for example, 01, 0101, 00000 etc.).
The polymerization progresses in a container with volume $V$ under a well-mixed (homogeneous) condition. Only the two monomer species flow into the container from outside at the same constant rate $\frac{1}{2}f V$, while all the molecules in the container diffuse out at rate $d$. 

For simplicity, only the ligation reaction between the polymers and monomers is considered here. The polymer has directionality, and the monomer can ligate from both the left and right sides of a polymer (e.g., $ 1 + 00 \rightarrow 100$ and  $00 + 1\rightarrow 001$ ). 
For simplicity and to focus on the kinetic aspect, all the polymer species of the same length are assumed to have the same chemical potential. Thus, in equilibrium, the concentrations of all the polymer species with the same length are equal.

A template polymer serving as a catalyst accelerates both the forward and backward reactions, without changing the equilibrium condition.
The catalytic reaction works if the template includes a subsequence of the bit inversion of the product. For example, the reaction $111 + 1  \rightarrow  1111$ is catalyzed by the template containing 0000 as a subsequence, i.e., $00000$, $10000$, or $00001$. 
For simplicity, only the longest polymers with the given length $L$, independent of their sequence, can serve as the template in ligation reactions.
Since all the ligation reactions do not change the total number of monomers, the ratio of monomer inflow to the outflow by the diffusion of the polymers determines the total number of monomers in the system, i.e., the total monomer concentration is given by $f/d$.

In summary, the chemical reactions are given as follows by denoting the polymer of the length $l$ with the sequence $s={\{0,1\}}^l$ as $A_{l,s}$:

\begin {equation*}
\begin {split}
\phi \overset {\frac{1}{2} f V} {\rightarrow} m, \quad 
A_ {l, s} + m \xrightarrow{\kappa_{l+1,sm}}  A_ {l + 1, sm}, \\
A_ {l, s} \overset {d} {\rightarrow} \phi, \quad 
m+ A_{l, s} \xrightarrow{\kappa_{l+1,ms}}  A_ {l + 1, ms}, \\
\end {split}
\end {equation*}
where $m$ is a monomer $0$ or $1$ and $sm$ and $ms$ are the sequences with $m$ added to the left side or right side of $s$, respectively. 
The polymerization can occur spontaneously at a rate $\epsilon$, which is set to be quite small.

Under the large volume limit, the concentration of polymer $A_{l,s}$, $x_{l,s}$ follows the deterministic rate equation.
\begin{equation}
\label{eq:rate}
\begin{split}
\dot{x}_{1,m} =&  \frac{1}{2} f -x_{1,m} \sum_{l',s'} (\kappa_{l'+1,s'm} + \kappa_{l'+1,ms'}) x_{l',s'} - d x_{1,m}, \\
\dot{x}_{l,s} =& \kappa_{l,s} x_{1,m} ( x_{l-1,s}^{(1)} + x_{l-1,s}^{(2)}  ) \\
 &- x_{l,s} \sum_{m'} x_{1,m'}( \kappa_{l+1,sm'}  + \kappa_{l+1,m's}  )- d x_{l,s},  \\
\dot{x}_{L,s} =& x_{1,m} x_{L,s} ( x_{L-1,s}^{(1)} + x_{L-1,s}^{(2)} ) - d x_{L,s}.
\end{split}
\end{equation} Here, $x_{l-1,s}^{(1)}$ and $x_{l-1,s}^{(2)}$ are the concentrations of the $(l-1)$-mer, which is a subsequence of $A_{l,s}$ on both sides \cite{about-dimer-reaction}, $ \kappa_{l,s} = \frac{1}{2} ( \epsilon + \sum_{(l',s') \in \mathcal{T}_{l+1,s}}  n_ {l',s'} / V  )$, and the $\mathcal{T}_{l,s}$ is the set of all template molecules that can catalyze the reactions to produce the polymer species $A_{l, s}$.
This equation has multiple attractors: As the pair of the longest polymer of a given sequence and its complementary pair catalyze the reaction to synthesize them, the concentrated population for each of the complementary pairs can be an attractor. It should be noted that since the complementary sequence works as a template for the polymerization, the complementary pair (e.g., 0010 and 1101) always coexists in an equal fraction.

\begin{figure}
\centering
\includegraphics[width=8.5cm]{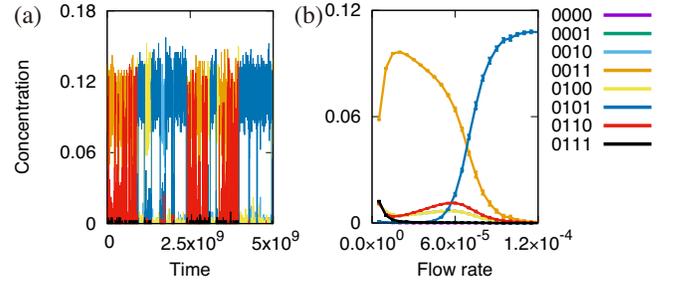}
\caption{(a) Time series of concentrations of all template species. $L=4, V=400, f/d = 1, \epsilon=10^{-4}$, $f=7.0 \times 10^{-5}$. (b) Average concentration of each template sequence in the stationary state. The horizontal axis is the flow rate in chemostat $f (=d)$.
The time intervals for average are varied from $10^{13}$ to $10^{14}$ depending on $f$. 
A sequence and its left-right reverse take the same average concentration over a long time, and the corresponding two lines are overlapped.
}
\label{fig:time_series}
\end{figure}

If the system size is finite, fluctuations around the rate Eq. \ref{eq:rate} induce switching among the attractors, and specific attractor(s) are selected dominantly. We investigated this selection by numerically solving the original stochastic reaction model\cite{gillespie1977exact}.
In Fig. \ref{fig:time_series}(a), the time series of the concentration of template species is plotted. A specific complementary pair of sequences is dominant for some time span, followed by switching to a state with a different dominant pair. As the concentrations of complementary sequences are equal, only one of them is plotted throughout the paper.

Next, the temporal average of the concentration of each template sequence was computed to examine its dependence on the flow rate, $f$ (Fig. \ref{fig:time_series}(b)), by fixing $f/d=1$ (hence, the total monomer concentration $f/d$ is kept constant). As shown, the concentration of each sequence, and in particular, that of the dominant species depend on $f$. Note that a sequence and its reverse form (e.g. 0010(1101) and 0100(1011) ) have the same fraction over a long time average due to symmetry, although at each instance, one of them is selected in the system. For $ L = 4 $, 0101 (and 1010) is selected for a large $f$, while 0011 is selected for a small $f$. For $L = 6$, as is shown in Supplemental Material\cite{supplement} Fig. S1(a), the dominant sequence changes as 010101 $\rightarrow$ 011001 $\rightarrow$ 001001 $\rightarrow$ 000100 with a decrease in $f$.

These changes in the dominant sequence against the decrease in the flow rate are interpreted as the increase in the sequence complexity. Here, a ``complex'' sequence implies that it contains more subsequences including their complementary ones. For example, 0101 includes only 01 and 10 dimers, while 0010 includes 00, 01, 10, and 11, and is thus more complex. In general, let $ \mathcal{C}_l (2k) $ be the set of template sequences that have $ 2 k $ species of $l$-mers as subsequences (note that the complementary sequence is included in the counting).
For example, 0101 belongs to $ \mathcal{C}_2(2) $ because it includes two dimer subsequences 01 and 10, and 0110 belongs to $\mathcal{C}_2(4)$ as it includes 01, 11, 10, and 00.
To study the selection of a complex sequence with $f$, we computed the sum of the average residence time at each attractor corresponding to the sequences belonging to $\mathcal{C}_l(2k)$. It is defined by $x_l(2k) =  \sum_ {s \in \mathcal{C}_l (2k)}  x_{L, s}$.
In Fig. S1(b) in the Supplemental Material\cite{supplement}, $x_l(2k)$ is plotted as a function of $f$. For  $L = 4$, the dominant concentrations change from $ x_2 (4) $ to $ x_2 (2) $, and for $ L = 6 $, they change in the order $ x_3 (8) $, $ x_3 (6) $, $ x_3 (4) $, $ x_3 (2) $ with an increase in $f$, implying that a more complex sequence is selected for a smaller $f$.

\begin{figure}
\centering
\includegraphics[width=6.8cm]{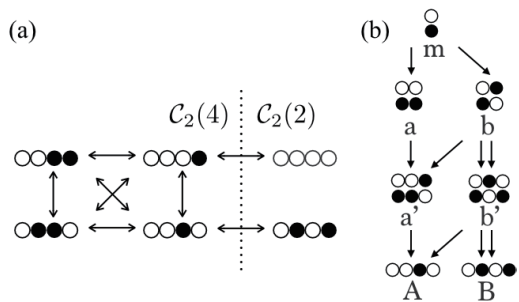}
\caption{(a) Transition network between attractors. Double head indicates transition between them and each sequence represents the one of a complementary pair that is dominant at each attractor.
(b) Schematic diagram representing the reaction pathway from monomers to each template $A$ and $B$. White and black circles represent $0$ and $1$ monomers, respectively. Arrows indicate ligation reaction of a shorter polymer with a monomer. Each reaction proceeds using $A$ or $B$ or both as a template. Detailed coefficients are in the main text. }
\label{fig:reaction_path}
\end{figure}

\begin{figure}
\includegraphics[width=8.5cm]{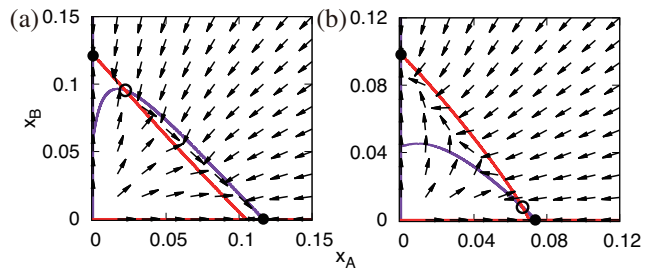}
\caption{Vector field of $x_A$ and $x_B$. Blue line is the nullcline of $\dot{x}_A=0$ and red is the nullcline of $\dot{x}_B=0$. 
The nullclines overlap $x_A$ line and $x_B$ line partially.  In both the figures, there are two stable fixed points ($\bullet$) and one unstable fixed point ($\circ$). 
Two stable fixed points are $A$-abundant and $B$-abundant attractors, respectively. As the value of $f$ increases from $f (=d) =1.0 \times 10^{-5}$(a) to $5.0 \times 10^{-4}$ (b), the unstable fixed point moves toward the $A$-abundant attractor.
}
\label{fig:vectorfield}
\end{figure}

To understand this change in the dominant sequence, we note that the residence time for each attractor, under noise due to the finiteness in the molecule number, is larger if the attraction toward it is stronger.
To understand this change in the dominant sequence, we first study the change from the sequence with $\mathcal{C}_2(4)$ to that with $\mathcal{C}_2(2)$.
Since the polymerization progresses from a dimer to a trimer, and then to a tetramer, the transition first occurs between tetramers that share the trimers, i.e., those with only a one-bit difference. The transition diagram is shown in Fig.  \ref{fig:reaction_path}(a), where the change from $\mathcal{C}_2(2)$ with 0101 to $\mathcal{C}_2(4)$ with a decrease in $f$ is mediated by the transition to sequence 1101 (0010). Hence, we first discuss the competition between the attractor with the dominance of the 0101 and 1010 pair (denoted as $B$) and that with the $0010$ and $1101$ pair (as $A$) by focusing only on these two pairs of templates, while neglecting other template polymers\cite{about-approximation}. By assuming that the number of shorter molecules changes faster than do the templates, the reduced rate equation is written only in terms of the concentrations of these templates, as follows:
\begin{eqnarray}
\label{eq:pol_A}
\dot{x}_A =& r_A(x_A, x_B) x_A &- d x_A,  \\
\label{eq:pol_B}
\dot{x}_B =& r_B(x_A, x_B) x_B &- d x_B.
\end{eqnarray}
Here, $r_A$ ($r_B$) is the synthesis rate of the corresponding template for each type $A$ ($B$) for a given $x_A$, $x_B$, respectively, whose form is obtained from Eq. S3, as shown in Supplemental Material\cite{supplement}. This rate equation has two fixed-point attractors, ($x_A = x_A^*, x_B= 0$) and ($x_A = 0, x_B= x_B^*$)\cite{about-spontaneoues}.
The vector field for $dx_A/dt$, $dx_B/dt$ is thus obtained as in Fig. \ref{fig:vectorfield}. Here, the attractor close to the unstable fixed point is each kicked out by noise.

It is shown straightforwardly that the attraction speed to attractor $A$ is given by $v_A\equiv r_B (x_A^*, 0) - r_A(x_A^*,0)$, from Eq. \ref{eq:pol_A}, \ref{eq:pol_B} and the eigenvalues of the Jacobian matrix of the dynamics, and that to $B$ is $v_B\equiv r_B(0,x_B^*) - r_A(0,x_B^*)$) (see Supplemental Material\cite{supplement}). As shown therein $v_B-v_A < 0$ for $f\sim 0$ and $ > 0$ for $f\gg1$, explaining the transition from $B$ to $A$ by noise with the decrease in $f$.
This flow rate dependence is also interpreted by the diversity in reaction pathways to replicate templates $A$ and $B$.
While only one reaction pathway exists for synthesizing $B$ from the monomers with double the speed because of symmetry, various pathways exist for synthesis of $A$ from the monomers because they have more types of subsequences.
For large $f(=d)$, double autocatalytic paths to $B$ work efficiently, while existence of diverse subsequences for the pathways to $A$ causes a larger loss of shorter polymers, leading to $r_B>r_A$. For small $f$, the concentrated use of the same dimer for $B$ results in its deficiency, and $r_A$ is larger since $A$ can use both dimers. 
(See Supplemental Material\cite{supplement} for analytic derivation of this dependence).

\begin{figure}
\centering
\includegraphics[height=4cm]{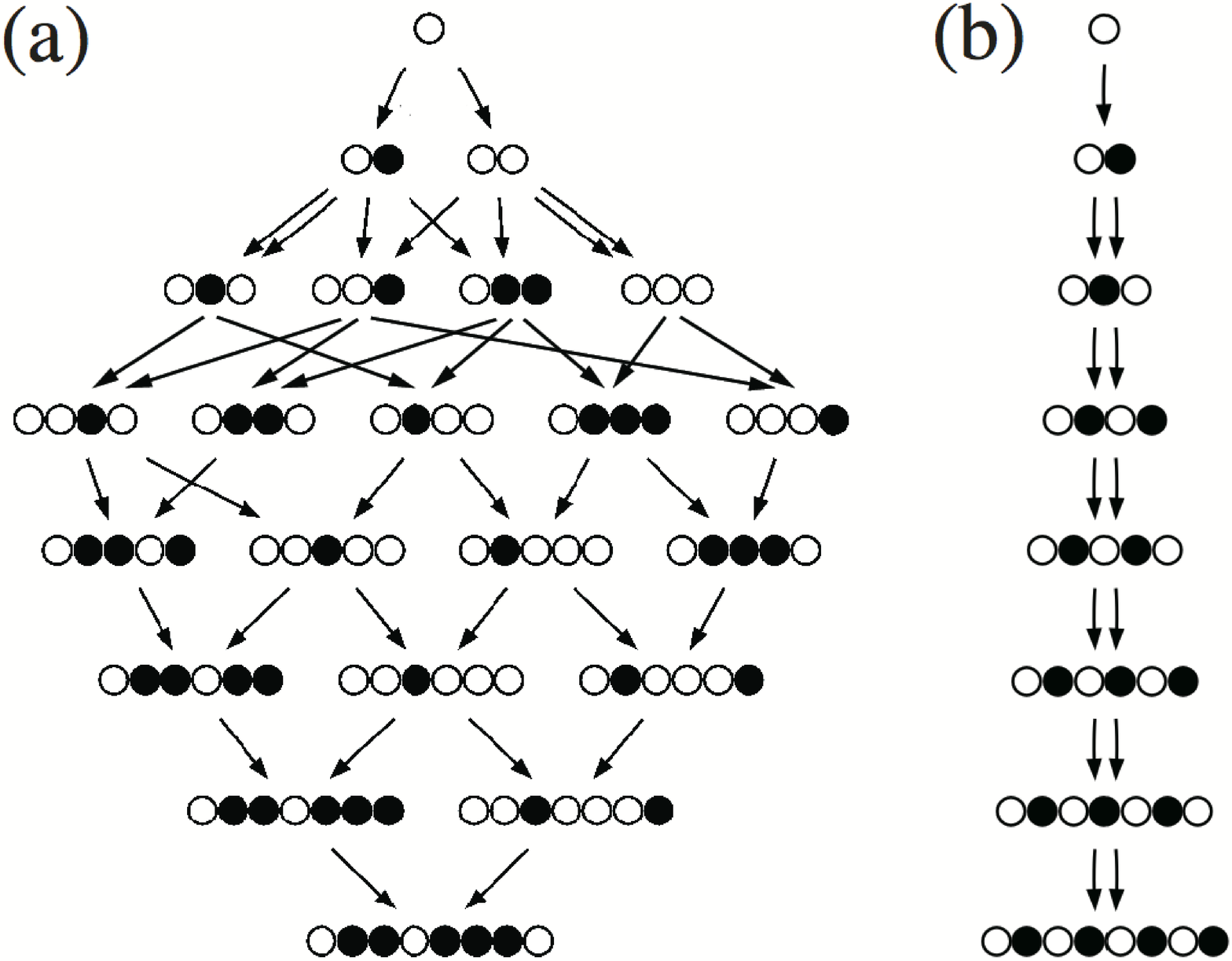}
\caption{ Reaction pathway to synthesize $8$-mer templates from monomers for (a) a complex sequence (01101110 $\in$ $\mathcal{C}_4(10)$) and (b) a simple sequence (01010101 $\in$ $\mathcal{C}_4(2)$). Black and white circles represent monomers 1 and 0, respectively. Each arrow indicates ligation reaction between a shorter polymer and a monomer, which are catalyzed by product templates.}
\label{fig:pathway}
\end{figure}

\begin{figure}
\includegraphics[width=8cm]{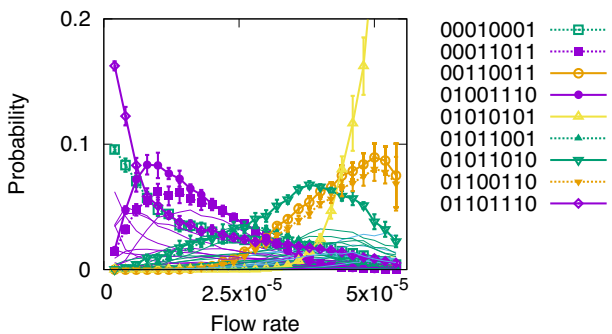}
\caption{
Residence probability of attractor at which a pair of sequences dominant. We set the length of the template $L$ as 8 and $\epsilon = 5.0 \times 10^{-6}$, $V=500$. The major sequences are plotted as thick lines. The probability is calculated as time average of $10^{14}\sim10^{15}$ simulation time. 
One of a reverse pair (e.g. 01101110 and 01110110) is omitted here. The complexity of the dominant sequence decreases with decreasing flow rate: The dominant sequence changes from 01010101 ($\mathcal{C}_4(2)$), 00110011 ($\mathcal{C}_4(4)$), 01011010 ($\mathcal{C}_4(6)$),  to 01001110 or 01101110 ($\mathcal{C}_4(10)$) with a decrease in flow rate.}
\label{fig:longer}
\end{figure}

The above argument holds for all the template sequences belonging to $\mathcal{C}_2(2)$ and $\mathcal{C}_2(4)$. Therefore, the total concentration of the sequence $x_2(2)$($x_2(4)$) is large when $f$ is large(small).
Furthermore, it is 
valid for the $L$-mer template sequence $A$ belonging to $ \mathcal{C}_l (2k+2) $ and $B$ belonging to $\mathcal{C}_l (2k)$ sharing subsequences in part. Here again, the replication speeds of $A$ and $B$, i.e., $r_A$ and $r_B$, determine the attraction speed to the fixed points $A$ and $B$, and the attractor $A$ with complex sequence is dominant for a low $f$. In Fig. \ref{fig:pathway}, the ligation reactions from the monomer to $8$-mer templates are drawn. The most complex sequence with $\mathcal{C}_4(10)$ has a variety of pathways, as compared with the simple sequence with $\mathcal{C}_4(2)$. For a low $f$, the sequence with $\mathcal{C}_4(10)$ is expected to be selected, while for a high $f$, that with $\mathcal{C}_4(2)$ is selected. The fraction of each sequence obtained from direct simulation is plotted in Fig. \ref{fig:longer} as a function of $f$.

Since direct simulation of cases with larger $L$ is time-consuming, we studied a reduced model, where only the transition between the attractors with a certain dominant template was considered, with the transition probability determined by the difference between the replication rates of the two templates (see Supplemental Material\cite{supplement} for details). A complex sequence with a large number of subsequences was selected as the flow rate decreased.

We have confirmed the generality of the present result, i.e., successive increase in the complexity of dominant sequence with the decrease in the flow rate, for the following cases (see Supplemental Material\cite{supplement}):
 (i) inclusion of variation in the catalytic activity by each template, if it is not too large
(ii) change in the rules for catalytic strength (iii) inclusion of the ligation between polymers longer than monomers
(iv) difference in the concentrations of monomers (0,1): In the study here, complementary polymers take the same concentrations, irrespective of the monomer concentrations, and thus, the bias in monomer cannot select one of the complementary molecules. 
(v) increase in spontaneous ligation rate $\epsilon$.
Finally, in the study presented here, the number of monomer types is only 2, in contrast to 4 in RNAs and 20 amino acids in  proteins.
Still, the increase in polymer complexity by the diversity of reaction pathways holds for a larger number of monomer types (see also \cite{tkachenko2017onset}).

To sum up, we have shown that in a polymerization process with template-based catalytic reactions, the dominant sequence selected depends on the flow rate of the monomers, and its complexity increases with the decrease in the rate. The selection is based on two basic mechanisms; (I) the attractor selection of a higher growth rate due to stochasticity in reactions, and (II) preference of a complex sequence to have a variety of pathways to avoid the deficiency of sub-part shorter polymers.
 
(I) The finiteness of the system size provides stochasticity (noise) in reaction, which gives rise to switches among attractors with different dominant sequences. Attractors with a higher growth rate have larger stability against noise (see for example Fig. 3), and are selected dominantly under the presence of noise. If the noise strength is too small, no transitions occur among the attractors, whereas if it is too large, no specific sequence dominates\cite{about-spontaneoues_reaction} (see also \cite{kashiwagi2006adaptive, furusawa2008generic} for attractor selection of a higher cellular growth, and \cite{Matsubara2016, jafarpour2015noise, kaneko2002kinetic} for relevance of noise to chemical evolution).
 
(II) When the flow rate is limited, the polymerization of simple (say 01-periodic) sequence suffers from the deficiency of subsequence polymers of shorter length.
The polymerization of a complex sequence of diverse pathways can keep a variety of subsequence polymers (see also \cite{kamimura2016negative} for an increase in component diversity in the protocell under low nutrient flow), while under a larger $d$ value, the decomposition of many components is disadvantageous for the growth.
 
This variety of subsequence leads to the definition of the complexity in sequence we adopted here\cite{about-complexity}. Of course, this is one possible definition, whereas a related definition is adopted to compare the subsequences in the genome in bioinformatics \cite{trifonov1990making,popov1996linguistic,  orlov2004complexity, shannon2001mathematical, kolmogorov1963tables, lloyd1988complexity}. The relationship between sequence complexity and structural or functional property of polymers has been discussed in several studies \cite{gabrielian1999sequence}. In our study, such a complex sequence is kinetically selected in the low flow rate region.

The results presented here will provide a novel perspective on kinetic conditions for the origin of life. Although a nonequilibrium condition is needed for life, our result implies that an excessively strong nonequilibrium condition will damage the sequence complexity. For the emergence and maintenance of prebiotic autocatalytic systems, optimal nonequilibrium conditions are required. Although it is difficult to estimate the actual flow rate for the appearance of complex sequences, the flux rate dependence of the complexity of synthesized polymers will be important to uncover the possible condition for the origin of life and for the laboratory construction of prebiotic self-replication systems.

\begin{acknowledgments}
We thank  Tetsuhiro S. Hatakeyama, Nobuto Takeuchi,  Atsushi Kamimura, and Nen Saito for the fruitful discussions. This research is partially supported by Grant-in-Aid for Scientific Research (S) (15H05746 [to K.K]) from the Japan Society for the Promotion of Science (JSPS) and JSPS KAKENHI Grant Number 17J07169 [to Y.J.M].
\end{acknowledgments}

\clearpage
\widetext
\setcounter{equation}{0}
\def\theequation{S\arabic{equation}}
\setcounter{figure}{0}
\def\thefigure{S\arabic{figure}}
\appendix

\section{Stationary distribution of the case with hexamer}
We calculated the average concentration of the template species following the same procedure as that for the tetramer case shown in Fig. 1(b), in the main text. The result showed a stepwise switch of the dominant sequence species. Note that as the flow rate reaches zero, the system approaches thermal equilibrium, where the frequency of each template sequence tends to be uniform (by considering also the backward reaction). The number of sequences that belong to $\mathcal{C}_3(6)$ is the largest among all the hexamers, and therefore, in Fig. S1(b), the frequency $x_3(6)$ increases as the flow rate approaches zero.

\begin{figure}[H]
\centering
\includegraphics[width=16cm]{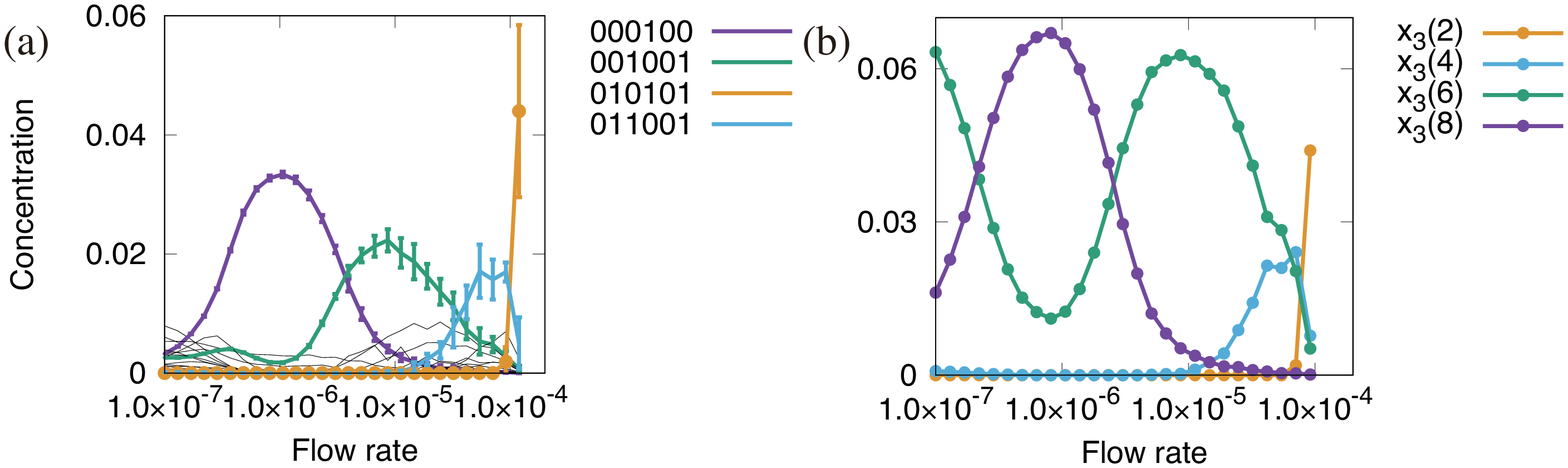}
\caption{ 
(a)Flow rate dependence of the average concentration of hexamer templates in stationary state.
The horizontal axis is the flow rate in chemostat $f (=d)$. 
We set $L=6, V=500, \epsilon=5.0 \times 10^{-6}$. The most dominant sequence is switched stepwise with the decrease in flow rate, as $010101 \rightarrow 011001 \rightarrow 001001 \rightarrow 001000$, which correspond to $\mathcal{C}_3(2)$, $\mathcal{C}_3(4)$, $\mathcal{C}_3(6)$, and $\mathcal{C}_3(8)$, respectively. For the flow rate $10^{-7}$, the distribution approaches that of thermal equilibrium. 
}
\label{fig:hexamer}
\end{figure}

\section{Detailed analysis in case of tetramer templates}

\subsection{Derivation of dynamics of two templates}
Considering the switch between the dominant sequences of 0010(1101) and 0101(1010), as mentioned in the main text, the rate equation is reduced to the concentrations of these templates and the corresponding trimers, dimers, and monomers.
\begin{equation*}
\begin{split}
\dot{x}_a &= \frac{1}{2} x_m^2 x_A - \frac{1}{2} x_m x_a x_A - d x_a \\
\dot{x}_b &= \frac{1}{2} x_m^2 ( 2 x_A + 2 x_B) - \frac{1}{2} x_m x_b ( 3 x_A + 4 x_B) - d x_b \\
\dot{x}_{a'} &= \frac{1}{2} x_m ( x_a + x_b ) x_A -  \frac{1}{2} x_m x_{a'} x_A - d x_{a'} \\
\dot{x}_{b'} &= \frac{1}{2} x_m 2 x_b (  x_A + 2 x_B) - \frac{1}{2} x_m ( x_A + 2 x_B) x_{b'} - d x_{b'} \\
\end{split}
\end{equation*}
\begin{eqnarray}
\label{eq:pol_A}
\dot{x}_A =& \frac{1}{2} x_m (x_{a'} + x_{b'}) x_A &- d x_A  \\
\label{eq:pol_B}
\dot{x}_B =&  x_m x_{b'} x_B &- d x_B
\end{eqnarray}

Here, $ x_m $ represents the concentration of 0 or 1, $ x_a $ is 00 or 11, $ x_b $ is 01 or 10, $ x_ {a'} $ is 001 or 110, $ x_ {b'} $ is 010 or 101 , $x_A $ is 0010 or 1101, and $ x_B $ is 0101 or 1010. Note that the respective complementary pair has the same concentration. 
The reaction coefficient in the above equations is proportional to the number of corresponding reactions. For example, 00 ($a$) is produced in a single reaction 0 + 0 + 0010, and is consumed in the reaction 00 + 1 + 0010 by the template 0010 ($A$). On the other hand, 01 ($b$) is produced in four reactions 0 + 1 + 0010, 0 + 1 + 1101, 0 + 1 + 0101, and 0 + 1 + 1010 (see main text Fig. 2(b)), where $B$ and $A$ are used in
 two reactions, respectively, and consumed in reactions 0 + 01 + 0010, 01 + 0 + 0010, 1 + 01 + 1101, 01 + 0 + 0101, 01 + 0 + 1010, 1 + 01 + 0101, and 1 + 01 + 1010, with three reactions using $A$ and four reactions using $B$.

The concentrations of $a, b, a', b'$ and $m$ are adiabatically eliminated, and two-variable dynamics just for $x_A$ and $x_B$ are obtained as Eq. \ref{eq:pol_A} and \ref{eq:pol_B} with
\begin{equation}
\label{eq:pol_conc}
\begin{split}
&{x_a} = \frac{x_A } {x_{m} x_A + 2 d } x_{m}^2  , \quad {x_b} = \frac{( 2 x_A + 2 x_B)}{  x_{m} ( 3 x_A + 4 x_B) + 2 d} x_{m}^2 , \\
&{x_{a'}} =  \frac{ x_{m}  x_A ( x_a + x_b )}{ x_m  x_A + 2 d }  , \quad {x_{b'}} = \frac{ x_{m} ( x_A + 2 x_B) (x_b+x_b)} { x_{m} ( x_A + 2 x_B) + 2 d }.
\end{split}
\end{equation}

Using them, the explicit form of $r_A$ and $r_B$ can be written as 
\begin{equation}
\begin{split}
r_A  &= \frac{1}{2}x_m^4 \left(   \frac{ x_A }{ x_m  x_A + 2 d } \left( \frac{x_A } {x_{m} x_A + 2 d }  +\frac{( 2 x_A + 2 x_B)}{  x_{m} ( 3 x_A + 4 x_B) +2  d}  \right)   + 2  \frac{  ( x_A + 2 x_B) } { x_{m} ( x_A + 2 x_B) + 2 d }  \frac{( 2 x_A + 2 x_B)}{  x_{m} ( 3 x_A + 4 x_B) + 2 d}      \right),\\
r_B  &= 2 x_m^4 \frac{  ( x_A + 2 x_B) } { x_{m} ( x_A + 2 x_B) + 2 d }  \frac{( 2 x_A + 2 x_B)}{  x_{m} ( 3 x_A + 4 x_B) + 2 d} .
\end{split}
\end{equation}
(Note that $x_m$ is also the function of $x_A$ and $x_B$ but cannot be solved explicitly.)

\subsection{Derivation of strength of attraction near the fixed point}

The Jacobian of the rate equation at a fixed point corresponding to the $A$-abundant attractor is given by
\begin{equation}
 J_A = 
 \begin{pmatrix}
\frac{ \partial r_A }{ \partial x_A} (x_A^*, 0)  x_A^* & \frac{ \partial r_A }{ \partial x_A} (x_A^*,0)  x_A^* \\ 
0 & r_B(x_A^*, 0) -d
 \end{pmatrix}.
 \end{equation}
Here, $r_A$ and $r_B$ are the replication rates of $A$ and $B$, respectively, and $ r_A (x_A,x_B) =  \frac{1}{2} x_m (x_{a'} + x_{b'}) $ and $r_B (x_A,x_B) = x_m x_{b'} $. 
The eigenvalues of this Jacobian are $\frac{ \partial r_A }{ \partial x_A} (x_A^*, 0)  x_A^*$ and $r_B (x_A^*, 0) -d$, and the  maximum eigenvalue of the Jacobian  is given by $r_B (x_A^*, 0) - r_A(x_A^*,0)$, considering that $r_A(x_A*, x_B^*) -d =0$. Therefore, the replication rates of $A$ and $B$, i.e., $r_A$ and $r_B$, determine the maximal eigenvalue of the Jacobian.  

\subsection{Flow rate dependence of concentration of shorter polymers and relation to reaction pathways}

For a large $f(= d)$, $x_a = x_A x_m^2/(2d)$ and $x_b = (2 x_A + 2 x_B) x_m^2 /(2d)$, so that $x_a < x_b$. 
The subsequence $b$ is produced to a greater extent because $b$ is shared by both $A$ and $B$, while the consumption of $b$ is irrelevant in the fast dilution case (in Eq. \ref{eq:pol_conc}, the dilution term $d$ in the denominator is much larger than the other term). 
If the supply of dimers is fast, the replication speed of the simple sequence $B$ is higher, as the reactions are concentrated on those using $b$.
Using this, the replication rates $r_A$ and $r_B$ become
\begin{equation}
\begin{split}
&r_A ( x_A, x_B) = \frac{x_m^4}{4d^2} (7 x_A^2 + 14 x_A x_B + 8 x_B^2 )  \\
&r_B ( x_A, x_B) =\frac{x_m^4}{d^2} (2 x_A^2 + 6 x_A x_B + 4 x_B^2 ).
\end{split}
\end{equation}
The attraction speed near the fixed point $A$ is given by $r_A ( x_A^* ,0 ) - r_B (x_A^*,0) (= -\frac{x_m^4}{4d^2}  {x_A^*}^2 ) < 0 $ and that near the fixed point $B$ is given by $r_B (0,x_B^*) - r_A ( 0 , x_B^* )  ( = 2  \frac{x_m^4}{d^2}  {x_B^*}^2  ) > 0 $. 

On the other hand, when $f(=m_0 d)$ is small, $ x_a = x_m$ and $x_b = \frac{2 x_A + 2 x_B}{3 x_A + 4 x_B} x_m $, so that $ x_a > x_b $. 
The replication rates are
\begin{equation}
\begin{split}
&r_A ( x_A, x_B) = x_m^2 ( 1 + 3 \frac{2 x_A + 2 x_B}{3 x_A+4 x_B} ) \\
&r_B ( x_A, x_B)  = x_m^2 ( 4  \frac{2 x_A + 2 x_B}{3 x_A+4 x_B}).
\end{split}
\end{equation}
When $f$ is small, the attraction speed toward the fixed point $A$ is given by $r_A ( x_A^* ,0 ) - r_B (x_A^*,0) (= \frac{8}{3} x_m^2 )> 0 $ and that toward the fixed point $B$ is given by $r_B (0,x_B^*) - r_A ( 0 , x_B^* ) ( = -\frac{1}{2} x_m^2) < 0 $.

In summary, the attraction to the fixed point $A$ is faster (slower) than that to $B$ when  $f$ is small (large).

\section{Reduced description of the model}

Since the original model requires excessive computational time, we examined the selection in a system size with larger $L$ using the reduced description. Here, the reduction scheme is as follows.

In the reduced model, we consider only the transition between the attractors, in which a certain pair of templates is dominant. During the transition, we assume that the number of templates other than the pair is zero. This is guaranteed when the spontaneous ligation rate, $\epsilon$, is small because a new template will appear only by a spontaneous reaction. For template length $L$, there are $2^{L-1}$ attractors. We estimate the transition probabilities, $T_{A \rightarrow B}$, between $A$-abundant and $B$-abundant attractors by noise.

First, we consider the dynamics of concentration of a pair of templates $A$ and $B$. By assuming that the dynamics of concentration of the shorter polymer is sufficiently faster than that for the template, the concentrations of polymers shorter than $L-1$ are adiabatically eliminated and determined as functions of the concentration of $A$ and $B$, i.e., $x_A$ and $x_B$, so that
\begin{equation}
\begin{split}
\dot{x}_A =& r_A(x_A, x_B) x_A - d x_A  \\
\dot{x}_B =& r_B(x_A, x_B) x_B - d x_B.
\end{split}
\end{equation}
This equation has two fixed points $x_A = x_A^*, x_B = 0 $ and $x_A=0, x_B = x_B^*$, where $r_A(x_A^*,0) = d$ and $r_B(0,x_B^*)=d$, corresponding to the $A$-abundant and $B$-abundant attractors, respectively. 

Then, the corresponding Langevin equations are given by
\begin{equation}
\begin{split}
\dot{x}_A = F_A(x_A, x_B)  + \sqrt{G_A(x_A, x_B)/V} \eta_A(t) \\
\dot{x}_B =F_B(x_A, x_B)  + \sqrt{G_B(x_A, x_B)/V} \eta_B(t),
\end{split}
\end{equation}
where $\eta_i(t)$ is the Gaussian white noise with $\langle \eta_i(t) \eta_j(t') \rangle = \delta_{ij} \delta(t-t')$ and $F_A  = r_A(x_A, x_B) x_A - d x_A $ , $G_A  = r_A(x_A, x_B) x_A + d x_A $.

We assume a one-dimensional slow manifold, in which the dynamics connecting the two attractors are restricted on the straight line between the fixed points $A$ and $B$. We also assume that the replication rate, $r_A$, and degradation rate, $d$, are constant along the line, so that $G_A =  2 x_A d $ and $G_B = 2 x_B d$.

Since we assume that the dynamics are restricted to the one-dimensional line, the probability that the system escapes from attractor $A$ and enters $B$ is calculated by the Kramers formulae, as follows (for example, \cite{gardiner1985stochastic_sm, van1992stochastic_sm}): 
\begin{equation}
T_{A \rightarrow B} \propto \frac{ 1} {G(x) } \exp(  2 V \int^x \frac{F(x')}{G(x')} dx'),
\end{equation}
where the integral $\int^x$ is carried over from fixed point A to point $x$ along the straight line connecting the fixed points $A$ and $B$, and $x$ is the value at which the integral takes the maximum value.

Using the approximation $x_A^* \sim x_B^*$, we can now obtain the transition probability explicitly.
$T_{A \rightarrow B} = \exp(- V \Delta_{A\rightarrow B})$, 
where $\Delta_{A\rightarrow B} = \frac{1}{\sqrt{2} d} \max_x \int_0^x \left( r_A - r_B  \right) dx' $. $\Delta_{A \rightarrow B}$ is considered to be the highest barrier in the quasi-potential of the dynamics of $x_A$ and $x_B$ along the line $x_A / x_A^* + x_B / x_B^* = 1$. 
(To calculate $r_A$ and $r_B$ explicitly, we also used $x_A^* \sim f/(2dL)$ by assuming that almost all the monomers ($f/d$) are included in the templates, and the free monomer concentration $x_m \sim \sqrt{f}$, as derived from Eq. 1 presented in the main text, by assuming that the concentrations of polymers other than the templates are sufficiently low.)

We calculated all the possible transitions between the attractors and constructed the transition matrix $T$. We assumed that transition occurs only between the pairs of $L$-mer templates that share the $(L-1)$-mer subsequences.  
By calculating the eigenvector with the maximum eigenvalue of the transition matrix $T$, we could obtain the stationary distribution of the model.

\begin{figure}[H]
\centering
\includegraphics[width=16cm]{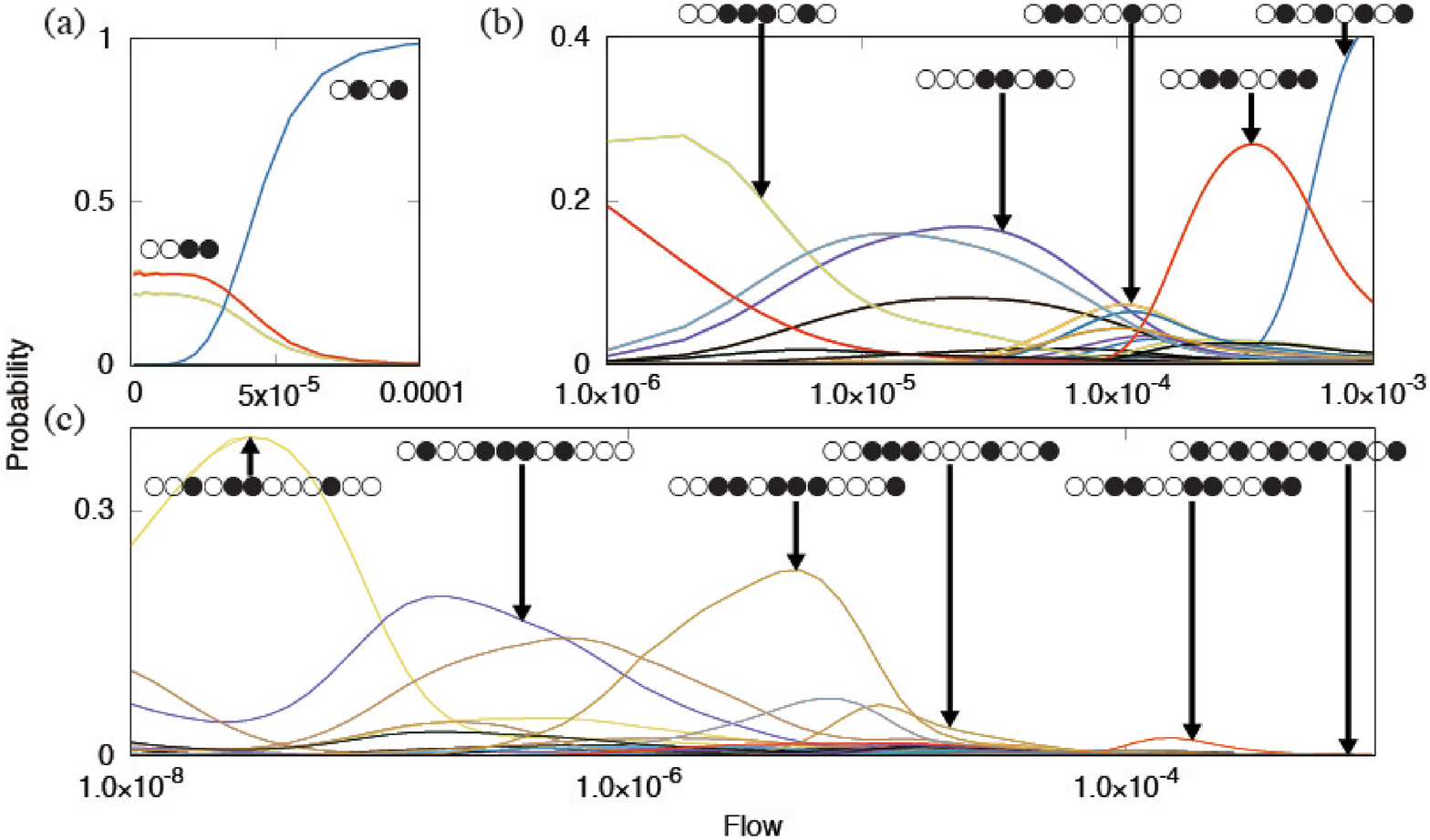}
\caption{Flow rate dependence of stationary probability distribution of attractors with dominance of a pair of templates. Systems with $L=4$(a), $L=8$(b), $L=12$(c) are plotted as a function of flow rate. $V/\sqrt{2} = 50$. Representative sequences are drawn in the figure (black and white circles indicate monomer type 0 and 1, respectively).}
\label{fig:msm}
\end{figure}

We calculated the residence probability of the attractors in the stationary distribution for $L=4, 8, 12$, as shown in Fig. \ref{fig:msm}, derived from the reduced model.

Although the obtained concentration of the template species differs quantitatively from that obtained by the original chemical reaction model for a small $L$, the stepwise increase in complexity in the dominant sequence with a decrease in flow rate was reproduced. In other words, at the highest flow rate, the periodic sequence (010101..., $\mathcal{C}_2(2)$) is selected. Here, we also examined the case with a longer $L$, e.g. $L=12$. As shown in Fig. S2(c), an increase in complexity was seen with a decrease in $f$.

\section{Generality of the results}

\subsection{(i) Variation in catalytic activity }

We tested the case in which the catalytic activity of each template $L$ for the sequence $s$ is not identical, given by $ k_{L,s}$, so that  $\kappa_{l,s}$ in the main text is replaced by $\frac{1}{2} ( \epsilon + \sum_{(L,s') \in \mathcal{T}_{l+1,s}} k_{L,s'} n_ {l',s'} / V)$.  
As an example, we studied the tetramer as the template, by changing the catalytic activity of the template 0101 (and 1010), which is the dominant sequence for fast flow, so that the activity for the sequence is given by $k$ while that for other sequences is set at unity.

\begin{figure}
\centering
\includegraphics[width=14cm]{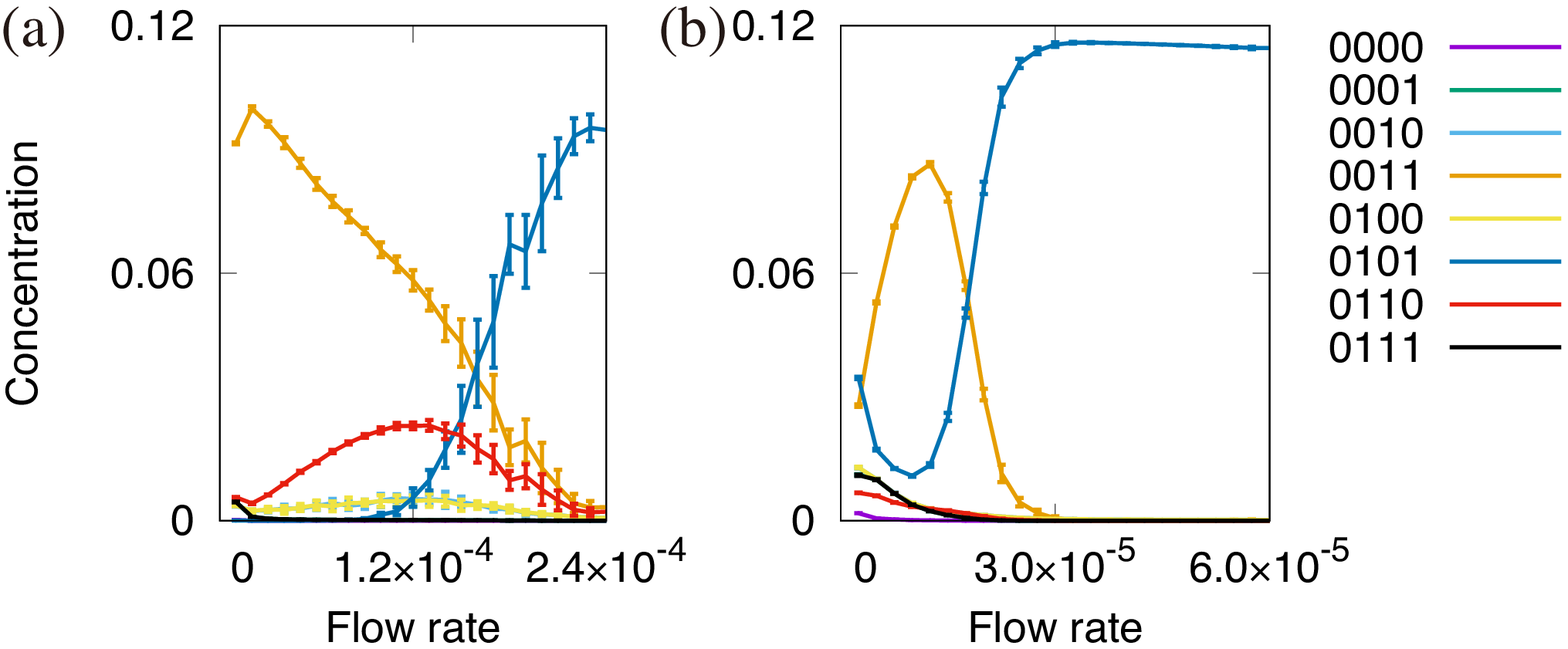}
\caption{  Average concentration of each template sequence in the  stationary state. The horizontal axis is the flow rate in chemostat $f (=d)$. $L=4, V=400, f/d = 1. \epsilon=10^{-4}$.  $k$ for the sequence 0101 (or 1010) is 0.9(a), 1.4(b) while $k_s$ for other sequences is kept at unity.
}
\label{fig:catalytic}
\end{figure}

In Fig. \ref{fig:catalytic}, we plotted the fraction of each sequence, as shown in Fig. 1b in the main text. The dominant templates change depending on the flow rate, whereas the transition flow rate and the frequency of dominant species changes so that the fraction of sequence 0101 is increased (decreased) with the increase in $k$.  If $k$ is too large (say $\sim 2$) ( too small (say $ \sim 0.5$) ), 0101 remains (is not) the dominant species, respectively, against the change in the flow rate. The competition between the difference in the catalytic activities and sequence diversity can be compared to the competition between the energy and entropy in equilibrium thermodynamics.

\subsection{(ii)Change in the ligation speed }

To examine the generality of the results in the main text, i.e., the increase in sequence complexity with the decrease in flow rate, we studied an alternative model in which the rules for catalytic strength were changed from that in the original model. 

In the alternative chemical reaction system, the ligation reaction rate of $l$-mer polymer with sequence $s$ ($A_{l,s}$)  $\kappa_{l,s}$ is changed as follows.
$$\kappa_{l,s} = \frac{1}{2} ( \epsilon + \sum_{(L,s') \in \mathcal{T}_{l+1,s}} m_{l,s, L,s'}  n_ {L,s'} / V),$$ where $m_{l,s, L,s'}$ is the number of $l$-mers with sequence $s$, which is contained in the template with sequence $s'$.
Under this rule, for example, the tetramer sequence $0101$ has two dimers $01$ as subsequences and catalyzes the reaction $0 +1 \rightarrow 01$ three times faster and $1+0 \rightarrow 10$ two times faster as compared to the catalytic strength in the original rule. 

We calculated the flow rate dependence of the average concentrations of templates in the stationary state (Fig. S3). Although the dominant sequence at each flow rate differs from that of the original rule (e.g. in the high flow rate case, not only 0101 but also 0000 is selected), a stepwise increase in the complexity of the dominant sequence with a decrease in flow rate is observed, as in the original model.

\begin{figure}
\centering
\includegraphics[width=16cm]{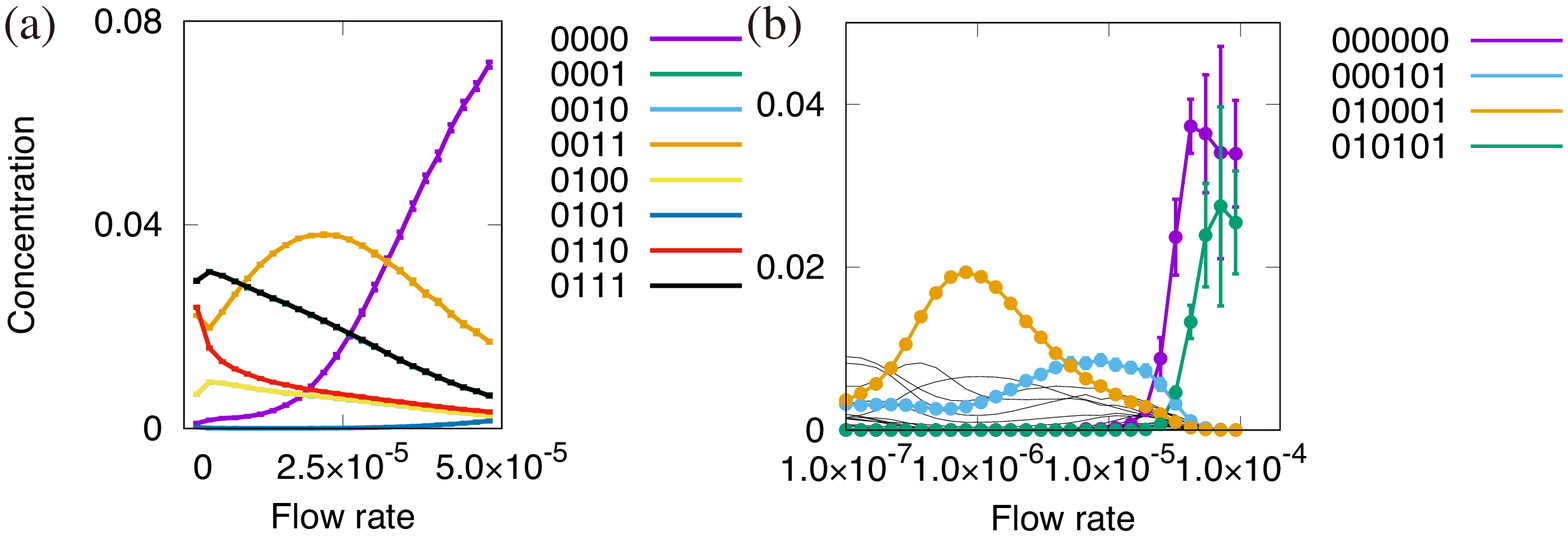}
\caption{ 
(a)Flow rate dependence of average concentrations of templates in the stationary state. The horizontal axis is the flow rate in chemostat $f (=d)$.  We set $L=4, V=200, \epsilon=1.0 \times 10^{-5}$ (a) and $L=6 , V=250, \epsilon=5.0 \times 10^{-6}$(b). The most dominant sequence is switched stepwise with the decrease in flow rate as $000000 \text{ or } 010101  \rightarrow 000101 \rightarrow 010001$, corresponding to $\mathcal{C}_3(2)$, $\mathcal{C}_3(6)$, and $\mathcal{C}_3(8)$, respectively. 
}
\label{fig:alter}
\end{figure}

\subsection{(iii)Ligation between polymers beyond the monomer}

\begin{figure}
\centering
\includegraphics[width=16cm]{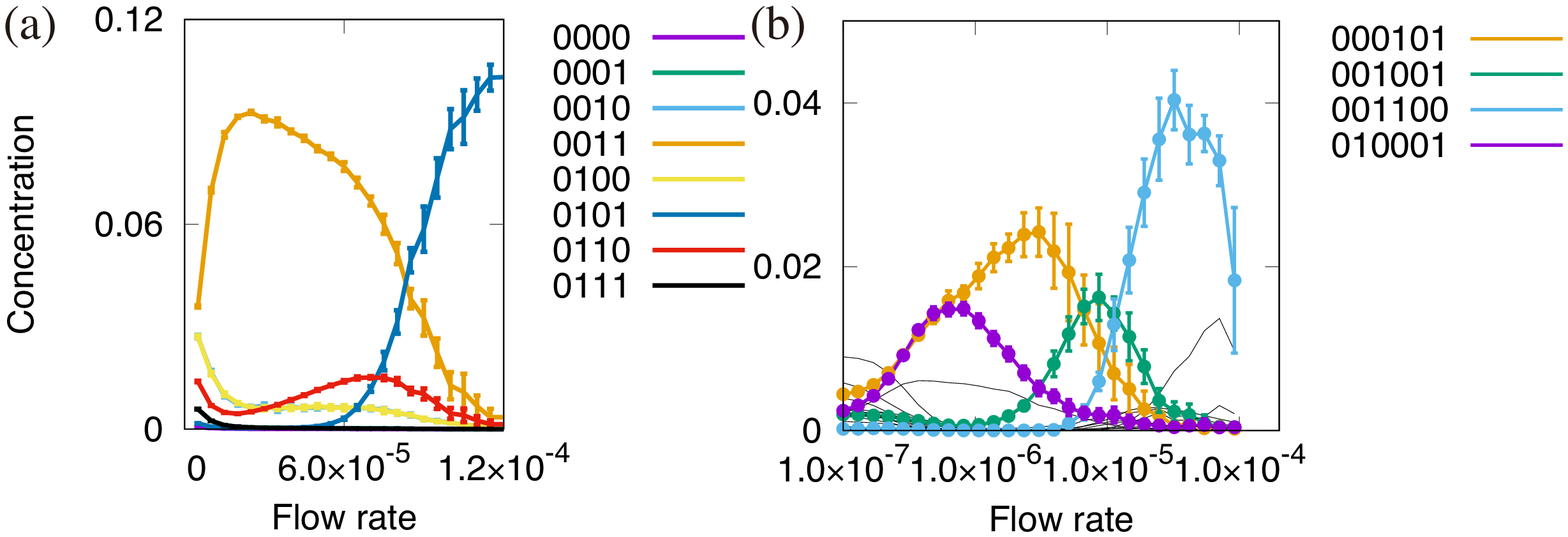}
\caption{ 
Flow rate dependence of average concentrations of templates in the stationary state. The horizontal axis is the flow rate in chemostat $f (=d)$.  We set $L=4, V=400, \epsilon=1.0 \times 10^{-4}$ (a) and $L=6 , V=250, \epsilon=1.0 \times 10^{-6}$.
The most dominant sequence is switched stepwise with the decrease in flow rate as $001100  \rightarrow 001001 \rightarrow 000101 \rightarrow 010001$, which correspond to $\mathcal{C}_3(4)$, $\mathcal{C}_3(6)$ ,$\mathcal{C}_3(6)$  and $\mathcal{C}_3(8)$, respectively. 
}
\label{fig:2p2}
\end{figure}

We also simulated another model in which there is not only ligation between a monomer and a polymer but also ligation between two polymers longer than the monomers (e.g. $01 + 01 \rightarrow 0101 $, $01 + 0101 \rightarrow 010101$), so that the ligation reaction in the original model is replaced by 
$$A_{l, s} + A_{l',s'} \xrightarrow{\kappa_{l+l',ss'}}  A_ {l + l', ss'}.
$$ 
 In this case also, the flow rate dependence of the complexity of the dominant sequence is observed, as in Fig.\ref{fig:2p2}.

\subsection{(iv) Difference between concentration of monomers 0 and 1}

\begin{figure}
\includegraphics[width=18cm]{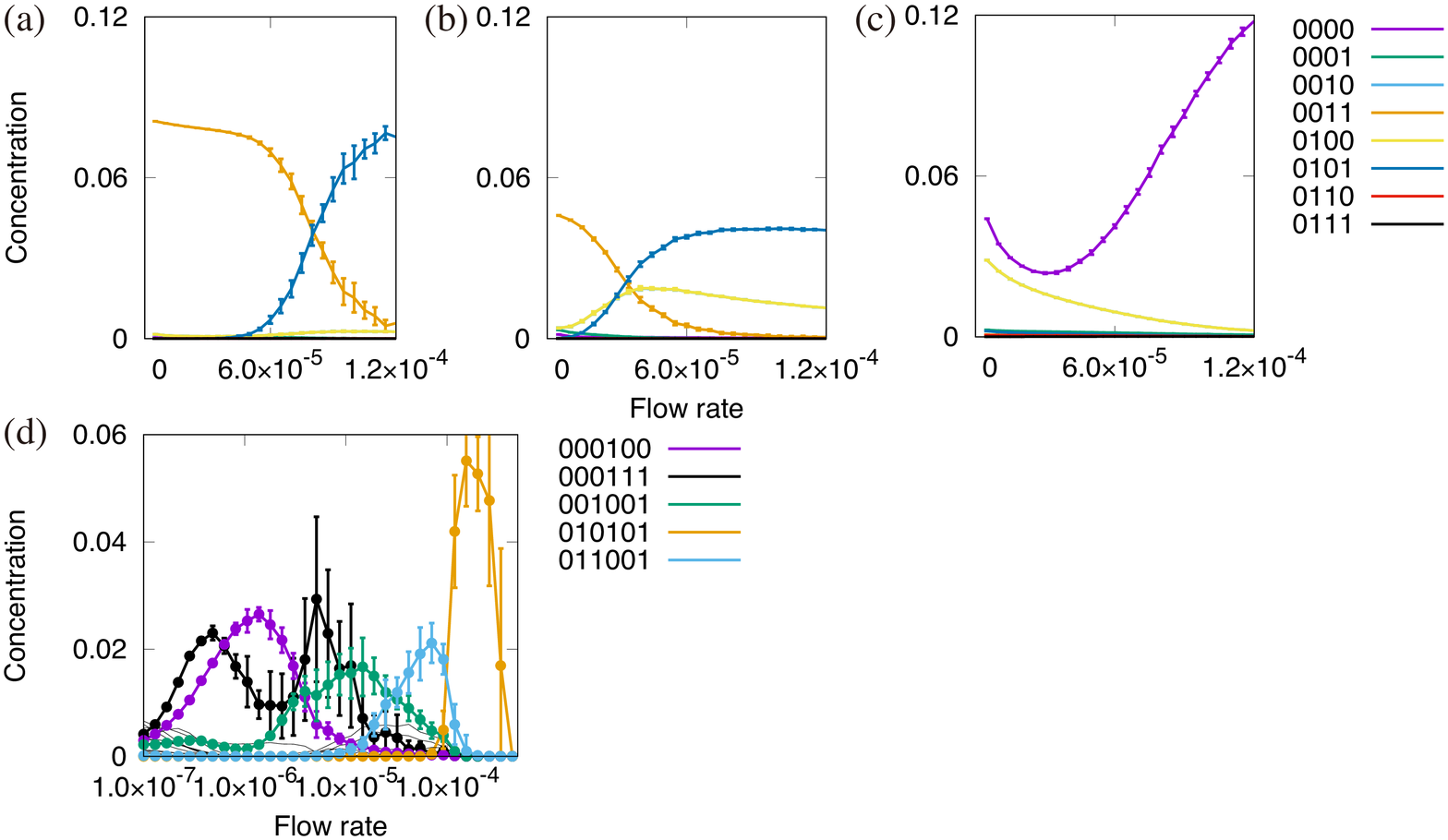}
\caption{Average concentration of each template sequence in the stationary state in the case of bias between monomer species. The horizontal axis is the flow rate in chemostat $f (=d)$. $L=4, V=400, f/d = 1. \epsilon=10^{-4}, f_0:f_1 = 0.6:0.4 (a) , 0.7:0.3 (b), 0.9:0.1 (c)$ and $L=6, f_0:f_1 = 0.525:0.475$ (d). The most dominant sequence is switched stepwise with the decrease in flow rate as $011001  \rightarrow 001001 \rightarrow 000111 \rightarrow 000100$, which correspond to $\mathcal{C}_3(4)$, $\mathcal{C}_3(6)$ ,$\mathcal{C}_3(6)$  and $\mathcal{C}_3(8)$, respectively. 
}
\label{fig:bias}
\end{figure}

In the main text, we assumed that the ratio between monomer species 0 and 1 was equal.
Here, we also calculated the case in which the supply rates of the two monomers are different, by varying the flow rate of monomers 0 given by $f_0$, and that of monomers 1 given by $f_1$, while keeping $f_0+f_1=1$ fixed.  The flow rate dependence of the averaged concentration of template species is shown  in Fig. \ref{fig:bias}.
If the bias between the two flow rates is not too large, qualitatively, the same result is obtained as in the main text Fig. 1b. This is because in our model, the templates replicate as a complementary pair between 0 and 1, in contrast with the model in ref.\cite{tkachenko2017onset_sm}. The same number of monomers (0 and 1) is consumed to make each complementary pair of polymers, so that complementary polymers (e.g., 0000 and 1111) take the same concentrations, irrespective, of the monomer concentrations. Thus, the bias in monomer cannot select one of the complementary molecules. 
As the bias increases, however, the synthesis of templates is slowed down because the minority monomer (i.e. monomer 1) is deficient. Then, the complementary pair hardly exists (i.e. an attractor at which a specific pair of templates exists is unstable) and the catalytic reaction stops. In this case, only the polymer consisting of the majority monomer (i.e. 0000) is synthesized by spontaneous (template independent) ligation reactions.

\subsection{ (v) Change in the spontaneous ligation rate $\epsilon$}
\begin{figure}
\centering
\includegraphics[width=14cm]{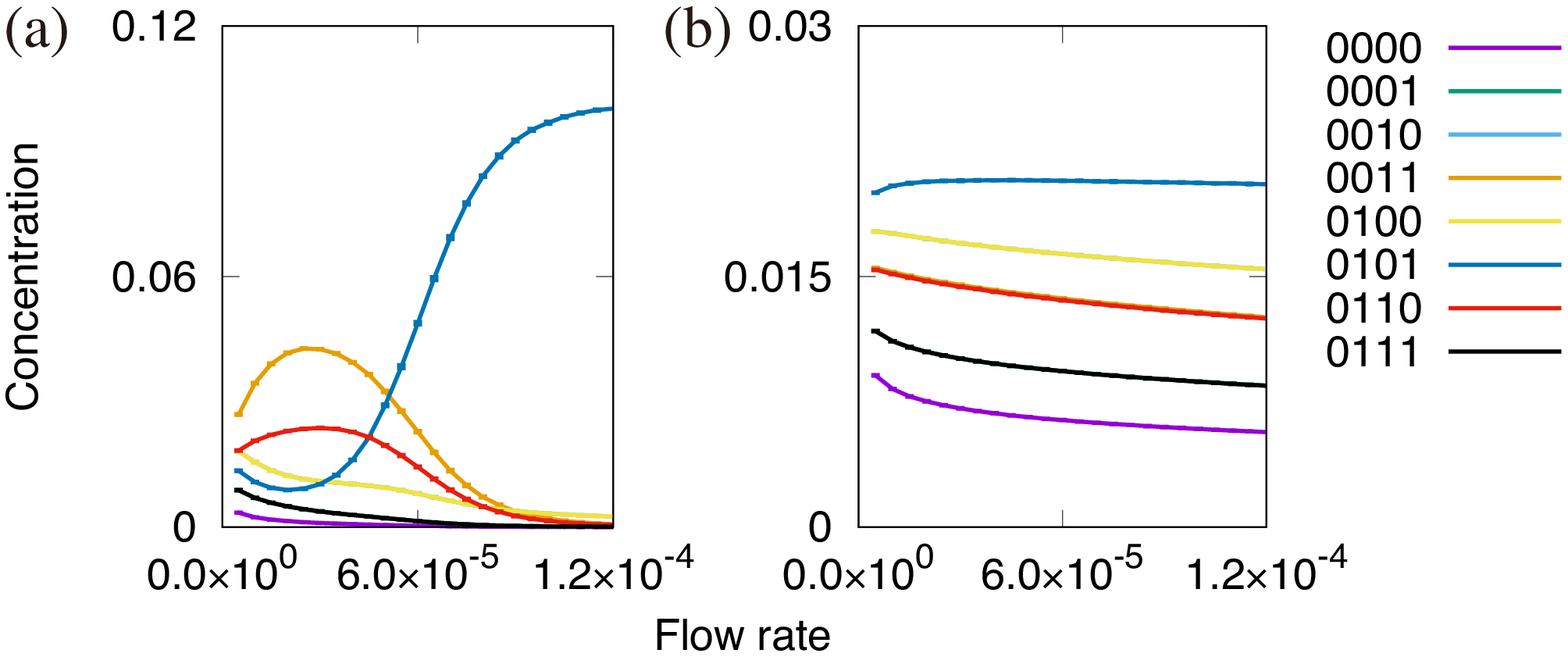}
\caption{ Flow rate dependence of average concentrations of templates in the stationary state. The horizontal axis is the flow rate in chemostat $f (=d)$. We set $L=4, V=400, \epsilon=1.0 \times 10^{-3}$ (a) and $\epsilon=1.0 \times 10^{-2}$ (b).
}
\label{fig:epsilon}
\end{figure}

We also studied dependence on the spontaneous ligation rate $\epsilon$. The flow rate dependence of the average concentration of template sequences is plotted in Fig. \ref{fig:epsilon}, as in the main text Fig. 1b. As shown, the result in Fig. 1b ( i.e. the dominance of sequence 0101 in the fast flow region, and 0011 in the slow flow region.) is robust against the increase in $\epsilon$ up to $ \sim 2.0\times 10^{-3}$. If $\epsilon$ is beyond this threshold, the attractor with a dominance of a certain template disappears and the template species just coexist (similar to the error catastrophe\cite{eigen1971selforganization_sm}), where the flow rate dependence is not observed.

\begin{thebibliography}{35}%
\makeatletter
\providecommand \@ifxundefined [1]{%
 \@ifx{#1\undefined}
}%
\providecommand \@ifnum [1]{%
 \ifnum #1\expandafter \@firstoftwo
 \else \expandafter \@secondoftwo
 \fi
}%
\providecommand \@ifx [1]{%
 \ifx #1\expandafter \@firstoftwo
 \else \expandafter \@secondoftwo
 \fi
}%
\providecommand \natexlab [1]{#1}%
\providecommand \enquote  [1]{``#1''}%
\providecommand \bibnamefont  [1]{#1}%
\providecommand \bibfnamefont [1]{#1}%
\providecommand \citenamefont [1]{#1}%
\providecommand \href@noop [0]{\@secondoftwo}%
\providecommand \href [0]{\begingroup \@sanitize@url \@href}%
\providecommand \@href[1]{\@@startlink{#1}\@@href}%
\providecommand \@@href[1]{\endgroup#1\@@endlink}%
\providecommand \@sanitize@url [0]{\catcode `\\12\catcode `\$12\catcode
  `\&12\catcode `\#12\catcode `\^12\catcode `\_12\catcode `\%12\relax}%
\providecommand \@@startlink[1]{}%
\providecommand \@@endlink[0]{}%
\providecommand \url  [0]{\begingroup\@sanitize@url \@url }%
\providecommand \@url [1]{\endgroup\@href {#1}{\urlprefix }}%
\providecommand \urlprefix  [0]{URL }%
\providecommand \Eprint [0]{\href }%
\providecommand \doibase [0]{http://dx.doi.org/}%
\providecommand \selectlanguage [0]{\@gobble}%
\providecommand \bibinfo  [0]{\@secondoftwo}%
\providecommand \bibfield  [0]{\@secondoftwo}%
\providecommand \translation [1]{[#1]}%
\providecommand \BibitemOpen [0]{}%
\providecommand \bibitemStop [0]{}%
\providecommand \bibitemNoStop [0]{.\EOS\space}%
\providecommand \EOS [0]{\spacefactor3000\relax}%
\providecommand \BibitemShut  [1]{\csname bibitem#1\endcsname}%
\let\auto@bib@innerbib\@empty
\bibitem [{\citenamefont {Schrodinger}(1943)}]{schrodinger1943life}%
  \BibitemOpen
  \bibfield  {author} {\bibinfo {author} {\bibfnamefont {E.}~\bibnamefont
  {Schr\"{o}dinger}},\ }\href@noop {} {\emph {\bibinfo {title} {What is Life?}}}\
  (\bibinfo  {publisher} {University Press, Cambridge},\ \bibinfo {year}
  {1943})\BibitemShut {NoStop}%
  \bibitem [{\citenamefont {von Kiedrowski}(1986)}]{von1986self}%
  \BibitemOpen
  \bibfield  {author} {\bibinfo {author} {\bibfnamefont {G.}~\bibnamefont {von
  Kiedrowski}},\ }\href@noop {} {\bibfield  {journal} {\bibinfo  {journal}
  {Angewandte Chemie International Edition in English}\ }\textbf {\bibinfo
  {volume} {25}},\ \bibinfo {pages} {932} (\bibinfo {year} {1986})}\BibitemShut
  {NoStop}%
\bibitem [{\citenamefont {Lincoln}\ and\ \citenamefont
  {Joyce}(2009)}]{lincoln2009self}%
  \BibitemOpen
  \bibfield  {author} {\bibinfo {author} {\bibfnamefont {T.~A.}\ \bibnamefont
  {Lincoln}}\ and\ \bibinfo {author} {\bibfnamefont {G.~F.}\ \bibnamefont
  {Joyce}},\ }\href@noop {} {\bibfield  {journal} {\bibinfo  {journal}
  {Science}\ }\textbf {\bibinfo {volume} {323}},\ \bibinfo {pages} {1229}
  (\bibinfo {year} {2009})}\BibitemShut {NoStop}%
\bibitem [{\citenamefont {Krammer}\ \emph {et~al.}(2012)\citenamefont
  {Krammer}, \citenamefont {M{\"o}ller},\ and\ \citenamefont
  {Braun}}]{krammer2012thermal}%
  \BibitemOpen
  \bibfield  {author} {\bibinfo {author} {\bibfnamefont {H.}~\bibnamefont
  {Krammer}}, \bibinfo {author} {\bibfnamefont {F.~M.}\ \bibnamefont
  {M{\"o}ller}}, \ and\ \bibinfo {author} {\bibfnamefont {D.}~\bibnamefont
  {Braun}},\ }\href@noop {} {\bibfield  {journal} {\bibinfo  {journal}
  {Phys. Rev. Lett. }\ }\textbf {\bibinfo {volume} {108}},\ \bibinfo
  {pages} {238104} (\bibinfo {year} {2012})}\BibitemShut {NoStop}%
\bibitem [{\citenamefont {Vaidya}\ \emph {et~al.}(2012)\citenamefont {Vaidya},
  \citenamefont {Manapat}, \citenamefont {Chen}, \citenamefont {Xulvi-Brunet},
  \citenamefont {Hayden},\ and\ \citenamefont
  {Lehman}}]{vaidya2012spontaneous}%
  \BibitemOpen
  \bibfield  {author} {\bibinfo {author} {\bibfnamefont {N.}~\bibnamefont
  {Vaidya}}, \bibinfo {author} {\bibfnamefont {M.~L.}\ \bibnamefont {Manapat}},
  \bibinfo {author} {\bibfnamefont {I.~A.}\ \bibnamefont {Chen}}, \bibinfo
  {author} {\bibfnamefont {R.}~\bibnamefont {Xulvi-Brunet}}, \bibinfo {author}
  {\bibfnamefont {E.~J.}\ \bibnamefont {Hayden}}, \ and\ \bibinfo {author}
  {\bibfnamefont {N.}~\bibnamefont {Lehman}},\ }\href@noop {} {\bibfield
  {journal} {\bibinfo  {journal} {Nature}\ }\textbf {\bibinfo {volume} {491}},\
  \bibinfo {pages} {72} (\bibinfo {year} {2012})}\BibitemShut {NoStop}%
\bibitem [{\citenamefont {Sadownik}\ \emph {et~al.}(2016)\citenamefont
  {Sadownik}, \citenamefont {Mattia}, \citenamefont {Nowak},\ and\
  \citenamefont {Otto}}]{sadownik2016diversification}%
  \BibitemOpen
  \bibfield  {author} {\bibinfo {author} {\bibfnamefont {J.~W.}\ \bibnamefont
  {Sadownik}}, \bibinfo {author} {\bibfnamefont {E.}~\bibnamefont {Mattia}},
  \bibinfo {author} {\bibfnamefont {P.}~\bibnamefont {Nowak}}, \ and\ \bibinfo
  {author} {\bibfnamefont {S.}~\bibnamefont {Otto}},\ }\href@noop {} {\bibfield
   {journal} {\bibinfo  {journal} {Nat. Chem.}\ }\textbf {\bibinfo
  {volume} {8}},\ \bibinfo {pages} {264} (\bibinfo {year} {2016})}\BibitemShut
  {NoStop}%
\bibitem [{\citenamefont {Carothers}\ \emph {et~al.}(2004)\citenamefont
  {Carothers}, \citenamefont {Oestreich}, \citenamefont {Davis},\ and\
  \citenamefont {Szostak}}]{carothers2004informational}%
  \BibitemOpen
  \bibfield  {author} {\bibinfo {author} {\bibfnamefont {J.~M.}\ \bibnamefont
  {Carothers}}, \bibinfo {author} {\bibfnamefont {S.~C.}\ \bibnamefont
  {Oestreich}}, \bibinfo {author} {\bibfnamefont {J.~H.}\ \bibnamefont
  {Davis}}, \ and\ \bibinfo {author} {\bibfnamefont {J.~W.}\ \bibnamefont
  {Szostak}},\ }\href@noop {} {\bibfield  {journal} {\bibinfo  {journal}
  {J. Am. Chem. Soc.}\ }\textbf {\bibinfo {volume}
  {126}},\ \bibinfo {pages} {5130} (\bibinfo {year} {2004})}\BibitemShut
  {NoStop}%
\bibitem [{\citenamefont {Derr}\ \emph {et~al.}(2012)\citenamefont {Derr},
  \citenamefont {Manapat}, \citenamefont {Rajamani}, \citenamefont {Leu},
  \citenamefont {Xulvi-Brunet}, \citenamefont {Joseph}, \citenamefont {Nowak},\
  and\ \citenamefont {Chen}}]{Derr2012}%
  \BibitemOpen
  \bibfield  {author} {\bibinfo {author} {\bibfnamefont {J.}~\bibnamefont
  {Derr}}, \bibinfo {author} {\bibfnamefont {M.~L.}\ \bibnamefont {Manapat}},
  \bibinfo {author} {\bibfnamefont {S.}~\bibnamefont {Rajamani}}, \bibinfo
  {author} {\bibfnamefont {K.}~\bibnamefont {Leu}}, \bibinfo {author}
  {\bibfnamefont {R.}~\bibnamefont {Xulvi-Brunet}}, \bibinfo {author}
  {\bibfnamefont {I.}~\bibnamefont {Joseph}}, \bibinfo {author} {\bibfnamefont
  {M.~A.}\ \bibnamefont {Nowak}}, \ and\ \bibinfo {author} {\bibfnamefont
  {I.~A.}\ \bibnamefont {Chen}},\ }\href {\doibase 10.1093/nar/gks065}
  {\bibfield  {journal} {\bibinfo  {journal} {Nucleic Acids Res.}\ }\textbf
  {\bibinfo {volume} {40}},\ \bibinfo {pages} {4711} (\bibinfo {year}
  {2012})}\BibitemShut {NoStop}%
\bibitem [{\citenamefont {Eigen}\ and\ \citenamefont
  {Schuster}(1977)}]{eigen1977principle}%
  \BibitemOpen
  \bibfield  {author} {\bibinfo {author} {\bibfnamefont {M.}~\bibnamefont
  {Eigen}}\ and\ \bibinfo {author} {\bibfnamefont {P.}~\bibnamefont
  {Schuster}},\ }\href@noop {} {\bibfield  {journal} {\bibinfo  {journal}
  {Naturwissenschaften}\ }\textbf {\bibinfo {volume} {64}},\ \bibinfo {pages}
  {541} (\bibinfo {year} {1977})}\BibitemShut {NoStop}%
\bibitem [{\citenamefont {Farmer}\ \emph {et~al.}(1986)\citenamefont {Farmer},
  \citenamefont {Kauffman},\ and\ \citenamefont
  {Packard}}]{farmer1986autocatalytic}%
  \BibitemOpen
  \bibfield  {author} {\bibinfo {author} {\bibfnamefont {J.~D.}\ \bibnamefont
  {Farmer}}, \bibinfo {author} {\bibfnamefont {S.~A.}\ \bibnamefont
  {Kauffman}}, \ and\ \bibinfo {author} {\bibfnamefont {N.~H.}\ \bibnamefont
  {Packard}},\ }\href@noop {} {\bibfield  {journal} {\bibinfo  {journal}
  {Physica D: Nonlinear Phenomena}\ }\textbf {\bibinfo {volume} {22}},\
  \bibinfo {pages} {50} (\bibinfo {year} {1986})}\BibitemShut {NoStop}%
\bibitem [{\citenamefont {Szab{\'o}}\ \emph {et~al.}(2002)\citenamefont
  {Szab{\'o}}, \citenamefont {Scheuring}, \citenamefont {Cz{\'a}r{\'a}n},\ and\
  \citenamefont {Szathm{\'a}ry}}]{szabo2002silico}%
  \BibitemOpen
  \bibfield  {author} {\bibinfo {author} {\bibfnamefont {P.}~\bibnamefont
  {Szab{\'o}}}, \bibinfo {author} {\bibfnamefont {I.}~\bibnamefont
  {Scheuring}}, \bibinfo {author} {\bibfnamefont {T.}~\bibnamefont
  {Cz{\'a}r{\'a}n}}, \ and\ \bibinfo {author} {\bibfnamefont {E.}~\bibnamefont
  {Szathm{\'a}ry}},\ }\href@noop {} {\bibfield  {journal} {\bibinfo  {journal}
  {Nature}\ }\textbf {\bibinfo {volume} {420}},\ \bibinfo {pages} {340}
  (\bibinfo {year} {2002})}\BibitemShut {NoStop}%
\bibitem [{\citenamefont {Kaneko}(2005)}]{kaneko2005recursive}%
  \BibitemOpen
  \bibfield  {author} {\bibinfo {author} {\bibfnamefont {K.}~\bibnamefont
  {Kaneko}},\ }\href@noop {} {\bibfield  {journal} {\bibinfo  {journal}
  {Adv. Chem. Phys.}\ }\textbf {\bibinfo {volume} {130}},\ \bibinfo
  {pages} {543} (\bibinfo {year} {2005})}\BibitemShut {NoStop}%
\bibitem [{\citenamefont {Giri}\ and\ \citenamefont
  {Jain}(2012)}]{giri2012origin}%
  \BibitemOpen
  \bibfield  {author} {\bibinfo {author} {\bibfnamefont {V.}~\bibnamefont
  {Giri}}\ and\ \bibinfo {author} {\bibfnamefont {S.}~\bibnamefont {Jain}},\
  }\href@noop {} {\bibfield  {journal} {\bibinfo  {journal} {PloS One}\
  }\textbf {\bibinfo {volume} {7}},\ \bibinfo {pages} {e29546} (\bibinfo {year}
  {2012})}\BibitemShut {NoStop}%
\bibitem [{\citenamefont {Matsubara}\ and\ \citenamefont
  {Kaneko}(2016)}]{Matsubara2016}%
  \BibitemOpen
  \bibfield  {author} {\bibinfo {author} {\bibfnamefont {Y.~J.}\ \bibnamefont
  {Matsubara}}\ and\ \bibinfo {author} {\bibfnamefont {K.}~\bibnamefont
  {Kaneko}},\ }\href {\doibase 10.1103/PhysRevE.93.032503} {\bibfield
  {journal} {\bibinfo  {journal} {Phys. Rev. E}\ }\textbf {\bibinfo
  {volume} {93}},\ \bibinfo {pages} {032503} (\bibinfo {year}
  {2016})}\BibitemShut {NoStop}%
\bibitem [{\citenamefont {Guseva}\ \emph {et~al.}(2017)\citenamefont {Guseva},
  \citenamefont {Zuckermann},\ and\ \citenamefont {Dill}}]{guseva2017foldamer}%
  \BibitemOpen
  \bibfield  {author} {\bibinfo {author} {\bibfnamefont {E.}~\bibnamefont
  {Guseva}}, \bibinfo {author} {\bibfnamefont {R.~N.}\ \bibnamefont
  {Zuckermann}}, \ and\ \bibinfo {author} {\bibfnamefont {K.~A.}\ \bibnamefont
  {Dill}},\ }\href@noop {} {\bibfield  {journal} {\bibinfo  {journal}
  {Proc. Natl. Acad. Sci. U. S. A.},\  \textbf {\bibinfo
  {volume} {114}},\  \bibinfo {pages}
  {E7460}} (\bibinfo {year} {2017})}\BibitemShut {NoStop}%
\bibitem [{\citenamefont {Anderson}(1983)}]{anderson1983suggested}%
  \BibitemOpen
  \bibfield  {author} {\bibinfo {author} {\bibfnamefont {P.~W.}\ \bibnamefont
  {Anderson}},\ }\href@noop {} {\bibfield  {journal} {\bibinfo  {journal}
  {Proc. Natl. Acad. Sci. U. S. A.}\ }\textbf {\bibinfo
  {volume} {80}},\ \bibinfo {pages} {3386} (\bibinfo {year}
  {1983})}\BibitemShut {NoStop}%
\bibitem [{\citenamefont {Stein}\ and\ \citenamefont
  {Anderson}(1984)}]{Stein1984}%
  \BibitemOpen
  \bibfield  {author} {\bibinfo {author} {\bibfnamefont {D.~L.}\ \bibnamefont
  {Stein}}\ and\ \bibinfo {author} {\bibfnamefont {P.~W.}\ \bibnamefont
  {Anderson}},\ }\href@noop {} {\bibfield  {journal} {\bibinfo  {journal}
  {Proc. Natl. Acad. Sci. U. S. A.}\ }\textbf {\bibinfo {volume} {81}},\ \bibinfo {pages} {1751}
  (\bibinfo {year} {1984})}\BibitemShut {NoStop}%
  \bibitem [{\citenamefont {Fernando}\ \emph {et~al.}(2007)\citenamefont
  {Fernando}, \citenamefont {Von~Kiedrowski},\ and\ \citenamefont
  {Szathm{\'a}ry}}]{fernando2007stochastic}%
  \BibitemOpen
  \bibfield  {author} {\bibinfo {author} {\bibfnamefont {C.}~\bibnamefont
  {Fernando}}, \bibinfo {author} {\bibfnamefont {G.}~\bibnamefont
  {Von~Kiedrowski}}, \ and\ \bibinfo {author} {\bibfnamefont {E.}~\bibnamefont
  {Szathm{\'a}ry}},\ }\href@noop {} {\bibfield  {journal} {\bibinfo  {journal}
  {Journal of Molecular Evolution}\ }\textbf {\bibinfo {volume} {64}},\
  \bibinfo {pages} {572} (\bibinfo {year} {2007})}\BibitemShut {NoStop}%
\bibitem [{\citenamefont {Nowak}\ and\ \citenamefont
  {Ohtsuki}(2008)}]{nowak2008prevolutionary}%
  \BibitemOpen
  \bibfield  {author} {\bibinfo {author} {\bibfnamefont {M.~A.}\ \bibnamefont
  {Nowak}}\ and\ \bibinfo {author} {\bibfnamefont {H.}~\bibnamefont
  {Ohtsuki}},\ }\href@noop {} {\bibfield  {journal} {\bibinfo  {journal}
  {Proc. Natl. Acad. Sci. U. S. A.}\ }\textbf {\bibinfo
  {volume} {105}},\ \bibinfo {pages} {14924} (\bibinfo {year}
  {2008})}\BibitemShut {NoStop}%
\bibitem [{\citenamefont {Ohtsuki}\ and\ \citenamefont
  {Nowak}(2009)}]{Ohtsuki2009}%
  \BibitemOpen
  \bibfield  {author} {\bibinfo {author} {\bibfnamefont {H.}~\bibnamefont
  {Ohtsuki}}\ and\ \bibinfo {author} {\bibfnamefont {M.}~\bibnamefont
  {Nowak}},\ }\href {\doibase 10.1098/rspb.2009.1136} {\bibfield  {journal}
  {\bibinfo  {journal} {Proc. R. Soc. B}\
  }\textbf {\bibinfo {volume} {276}},\ \bibinfo {pages} {3783} (\bibinfo {year}
  {2009})}\BibitemShut {NoStop}%
\bibitem [{\citenamefont {Obermayer}\ \emph {et~al.}(2011)\citenamefont
  {Obermayer}, \citenamefont {Krammer}, \citenamefont {Braun},\ and\
  \citenamefont {Gerland}}]{Obermayer2011}%
  \BibitemOpen
  \bibfield  {author} {\bibinfo {author} {\bibfnamefont {B.}~\bibnamefont
  {Obermayer}}, \bibinfo {author} {\bibfnamefont {H.}~\bibnamefont {Krammer}},
  \bibinfo {author} {\bibfnamefont {D.}~\bibnamefont {Braun}}, \ and\ \bibinfo
  {author} {\bibfnamefont {U.}~\bibnamefont {Gerland}},\ }\href {\doibase
  10.1103/PhysRevLett.107.018101} {\bibfield  {journal} {\bibinfo  {journal}
  {Phys. Rev. Lett.}\ }\textbf {\bibinfo {volume} {107}},\ \bibinfo
  {pages} {018101} (\bibinfo {year} {2011})}\BibitemShut {NoStop}%
\bibitem [{\citenamefont {Tanaka}\ \emph {et~al.}(2014)\citenamefont {Tanaka},
  \citenamefont {Fellermann},\ and\ \citenamefont {Rasmussen}}]{Tanaka2014}%
  \BibitemOpen
  \bibfield  {author} {\bibinfo {author} {\bibfnamefont {S.}~\bibnamefont
  {Tanaka}}, \bibinfo {author} {\bibfnamefont {H.}~\bibnamefont {Fellermann}},
  \ and\ \bibinfo {author} {\bibfnamefont {S.}~\bibnamefont {Rasmussen}},\
  }\href {\doibase 10.1209/0295-5075/107/28004} {\bibfield  {journal} {\bibinfo
   {journal} {EPL}\ }\textbf {\bibinfo {volume} {107}},\
  \bibinfo {pages} {28004} (\bibinfo {year} {2014})}\BibitemShut {NoStop}%
\bibitem [{\citenamefont {Tkachenko}\ and\ \citenamefont
  {Maslov}(2015)}]{Tkachenko2015}%
  \BibitemOpen
  \bibfield  {author} {\bibinfo {author} {\bibfnamefont {A.~V.}\ \bibnamefont
  {Tkachenko}}\ and\ \bibinfo {author} {\bibfnamefont {S.}~\bibnamefont
  {Maslov}},\ }\href {\doibase 10.1063/1.4922545} {\bibfield  {journal}
  {\bibinfo  {journal} {J. Chem. Phys.}\ }\textbf {\bibinfo
  {volume} {143}},\ \bibinfo {pages} {045102} (\bibinfo {year}
  {2015})}\BibitemShut {NoStop}%
  \bibitem [{\citenamefont {Kinsler}\ \emph {et~al.}(2017)\citenamefont
  {Kinsler}, \citenamefont {Sinai}, \citenamefont {Lee},\ and\ \citenamefont
  {Nowak}}]{kinsler2017prebiotic}%
  \BibitemOpen
  \bibfield  {author} {\bibinfo {author} {\bibfnamefont {G.}~\bibnamefont
  {Kinsler}}, \bibinfo {author} {\bibfnamefont {S.}~\bibnamefont {Sinai}},
  \bibinfo {author} {\bibfnamefont {N.~K.}\ \bibnamefont {Lee}}, \ and\
  \bibinfo {author} {\bibfnamefont {M.~A.}\ \bibnamefont {Nowak}},\ }\href@noop
  {} {\bibfield  {journal} {\bibinfo  {journal} {PloS one}\ }\textbf {\bibinfo
  {volume} {12}},\ \bibinfo {pages} {e0180208} (\bibinfo {year}
  {2017})}\BibitemShut {NoStop}%
\bibitem [{\citenamefont {He}\ \emph {et~al.}(2016)\citenamefont {He},
  \citenamefont {G{\'a}llego}, \citenamefont {Laughlin}, \citenamefont
  {Grover},\ and\ \citenamefont {Hud}}]{he2016viscous}%
  \BibitemOpen
  \bibfield  {author} {\bibinfo {author} {\bibfnamefont {C.}~\bibnamefont
  {He}}, \bibinfo {author} {\bibfnamefont {I.}~\bibnamefont {G{\'a}llego}},
  \bibinfo {author} {\bibfnamefont {B.}~\bibnamefont {Laughlin}}, \bibinfo
  {author} {\bibfnamefont {M.~A.}\ \bibnamefont {Grover}}, \ and\ \bibinfo
  {author} {\bibfnamefont {N.~V.}\ \bibnamefont {Hud}},\ }\href@noop {}
  {\bibfield  {journal} {\bibinfo  {journal} {Nat. Chem.}\ }\textbf {\bibinfo
  {volume} {9}},\ \bibinfo {pages} {318} (\bibinfo
  {year} {2017})}\BibitemShut {NoStop}%
  \bibitem [{\citenamefont {Toyabe}\ and\ \citenamefont
  {Braun}(2018)}]{toyabe2018templated}%
  \BibitemOpen
  \bibfield  {author} {\bibinfo {author} {\bibfnamefont {S.}~\bibnamefont
  {Toyabe}}\ and\ \bibinfo {author} {\bibfnamefont {D.}~\bibnamefont {Braun}},\
  }\href@noop {} {\bibfield  {journal} {\bibinfo  {journal} {arXiv preprint
  arXiv:1802.06544}\ } (\bibinfo {year} {2018})}\BibitemShut {NoStop}%
\bibitem [{\citenamefont {Novick}\ and\ \citenamefont
  {Szilard}(1950)}]{novick1950description}%
  \BibitemOpen
  \bibfield  {author} {\bibinfo {author} {\bibfnamefont {A.}~\bibnamefont
  {Novick}}\ and\ \bibinfo {author} {\bibfnamefont {L.}~\bibnamefont
  {Szilard}},\ }\href@noop {} {\bibfield  {journal} {\bibinfo  {journal}
  {Science}\ }\textbf {\bibinfo {volume} {112}},\ \bibinfo {pages} {715}
  (\bibinfo {year} {1950})}\BibitemShut {NoStop}%
\bibitem []{about-ligation-reaction}%
  \BibitemOpen
For $l=1$ the top and the bottom reactions are considered only once.
\bibitem []{about-backward-reaction}%
  \BibitemOpen
The reverse ligation reaction is neglected here, but
its inclusion does not alter the following results qualitatively,
as long as its rate is small.
\bibitem []{about-dimer-reaction}%
  \BibitemOpen
For $l=2$, eq. 1 should be read with $x_{1,s}^{(1)} = x_{1,m}$ and $x_{1,s}^{(2)} = 0$).
\bibitem [{\citenamefont {Gillespie}(1977)}]{gillespie1977exact}%
  \BibitemOpen
  \bibfield  {author} {\bibinfo {author} {\bibfnamefont {D.~T.}\ \bibnamefont
  {Gillespie}},\ }\href@noop {} {\bibfield  {journal} {\bibinfo  {journal} {
  J. Phys. Chem.}\ }\textbf {\bibinfo {volume} {81}},\ \bibinfo
  {pages} {2340} (\bibinfo {year} {1977})}\BibitemShut {NoStop}%
\bibitem []{supplement}%
  \BibitemOpen See Supplemental Material at ??? for figures
with additional information on the case with hexamer, detailed analysis in the case of tetramer, reduced description of model and examination of generality of the result, which include Refs. \cite{gardiner1985stochastic, van1992stochastic, eigen1971selforganization}.
\bibitem [{\citenamefont {Gardiner}(1985)}]{gardiner1985stochastic}%
  \BibitemOpen
  \bibfield  {author} {\bibinfo {author} {\bibfnamefont {C.}~\bibnamefont
  {Gardiner}},\ }\href@noop {} {\emph {\bibinfo {title} {Stochastic Methods}}}\
  (\bibinfo  {publisher} {Springer-Verlag, Berlin--Heidelberg--New
  York--Tokyo},\ \bibinfo {year} {1985})\BibitemShut {NoStop}%
\bibitem [{\citenamefont {Van~Kampen}(1992)}]{van1992stochastic}%
  \BibitemOpen
  \bibfield  {author} {\bibinfo {author} {\bibfnamefont {N.~G.}\ \bibnamefont
  {Van~Kampen}},\ }\href@noop {} {\emph {\bibinfo {title} {Stochastic Processes
  in Physics and Chemistry}}},\ Vol.~\bibinfo {volume} {1}\ (\bibinfo
  {publisher} {Elsevier},\ \bibinfo {year} {1992})\BibitemShut {NoStop}%
    \bibitem [{\citenamefont {Eigen}(1971)}]{eigen1971selforganization}%
  \BibitemOpen
  \bibfield  {author} {\bibinfo {author} {\bibfnamefont {M.}~\bibnamefont
  {Eigen}},\ }\href@noop {} {\bibfield  {journal} {\bibinfo  {journal}
  {Naturwissenschaften}\ }\textbf {\bibinfo {volume} {58}},\ \bibinfo {pages}
  {465} (\bibinfo {year} {1971})}\BibitemShut {NoStop}%
    \bibitem [{abo({\natexlab{b}})}]{about-approximation}%
  \BibitemOpen
This approximation is valid when the rate of spontaneous ligation is much smaller than the time scale of the transition dynamics.
    \bibitem [{abo({\natexlab{b}})}]{about-spontaneoues}%
  \BibitemOpen
To analyze the stability of each attractor, the 
(tiny) spontaneous ligation term is not necessary,
and it is neglected.
 \bibitem [{\citenamefont {Tkachenko}\ and\ \citenamefont
  {Maslov}(2017)}]{tkachenko2017onset}%
  \BibitemOpen
  \bibfield  {author} {\bibinfo {author} {\bibfnamefont {A.~V.}\ \bibnamefont
  {Tkachenko}}\ and\ \bibinfo {author} {\bibfnamefont {S.}~\bibnamefont
  {Maslov}},\ }\href@noop {} {\bibfield  {journal} {\bibinfo  {journal} {arXiv
  preprint arXiv:1710.06385}\ } (\bibinfo {year} {2017})}\BibitemShut {NoStop}%
    \bibitem [{abo({\natexlab{b}})}]{about-spontaneoues_reaction}%
  \BibitemOpen
 Spontaneous reaction that undergoes without templates plays the similar role as the noise: If it is too large, there emerges no dominant sequence, and if it is too small, there is no transition between the attractors\cite{supplement}.
\bibitem [{\citenamefont {Kashiwagi}\ \emph {et~al.}(2006)\citenamefont
  {Kashiwagi}, \citenamefont {Urabe}, \citenamefont {Kaneko},\ and\
  \citenamefont {Yomo}}]{kashiwagi2006adaptive}%
  \BibitemOpen
  \bibfield  {author} {\bibinfo {author} {\bibfnamefont {A.}~\bibnamefont
  {Kashiwagi}}, \bibinfo {author} {\bibfnamefont {I.}~\bibnamefont {Urabe}},
  \bibinfo {author} {\bibfnamefont {K.}~\bibnamefont {Kaneko}}, \ and\ \bibinfo
  {author} {\bibfnamefont {T.}~\bibnamefont {Yomo}},\ }\href@noop {} {\bibfield
   {journal} {\bibinfo  {journal} {PloS One}\ }\textbf {\bibinfo {volume}
  {1}},\ \bibinfo {pages} {e49} (\bibinfo {year} {2006})}\BibitemShut {NoStop}%
\bibitem [{\citenamefont {Furusawa}\ and\ \citenamefont
  {Kaneko}(2008)}]{furusawa2008generic}%
  \BibitemOpen
  \bibfield  {author} {\bibinfo {author} {\bibfnamefont {C.}~\bibnamefont
  {Furusawa}}\ and\ \bibinfo {author} {\bibfnamefont {K.}~\bibnamefont
  {Kaneko}},\ }\href@noop {} {\bibfield  {journal} {\bibinfo  {journal} {PLoS
  Comput. Biol.}\ }\textbf {\bibinfo {volume} {4}},\ \bibinfo {pages}
  {e3} (\bibinfo {year} {2008})}\BibitemShut {NoStop}%
\bibitem [{\citenamefont {Jafarpour}\ \emph {et~al.}(2015)\citenamefont
  {Jafarpour}, \citenamefont {Biancalani},\ and\ \citenamefont
  {Goldenfeld}}]{jafarpour2015noise}%
  \BibitemOpen
  \bibfield  {author} {\bibinfo {author} {\bibfnamefont {F.}~\bibnamefont
  {Jafarpour}}, \bibinfo {author} {\bibfnamefont {T.}~\bibnamefont
  {Biancalani}}, \ and\ \bibinfo {author} {\bibfnamefont {N.}~\bibnamefont
  {Goldenfeld}},\ }\href@noop {} {\bibfield  {journal} {\bibinfo  {journal}
  {Phys. Rev. Lett.}\ }\textbf {\bibinfo {volume} {115}},\ \bibinfo
  {pages} {158101} (\bibinfo {year} {2015})}\BibitemShut {NoStop}%
\bibitem [{\citenamefont {Kaneko}\ and\ \citenamefont
  {Yomo}(2002)}]{kaneko2002kinetic}%
  \BibitemOpen
  \bibfield  {author} {\bibinfo {author} {\bibfnamefont {K.}~\bibnamefont
  {Kaneko}}\ and\ \bibinfo {author} {\bibfnamefont {T.}~\bibnamefont {Yomo}},\
  }\href@noop {} {\bibfield  {journal} {\bibinfo  {journal} {J.
  Theor. Biol.}\ }\textbf {\bibinfo {volume} {214}},\ \bibinfo {pages}
  {563} (\bibinfo {year} {2002})}\BibitemShut {NoStop}%
  \bibitem [{\citenamefont {Kamimura}\ and\ \citenamefont
  {Kaneko}(2016)}]{kamimura2016negative}%
  \BibitemOpen
  \bibfield  {author} {\bibinfo {author} {\bibfnamefont {A.}~\bibnamefont
  {Kamimura}}\ and\ \bibinfo {author} {\bibfnamefont {K.}~\bibnamefont
  {Kaneko}},\ }\href@noop {} {\bibfield  {journal} {\bibinfo  {journal}
  {Phys. Rev. E}\ }\textbf {\bibinfo {volume} {93}},\ \bibinfo {pages}
  {062419} (\bibinfo {year} {2016})}\BibitemShut {NoStop}%
     \bibitem [{abo({\natexlab{b}})}]{about-complexity}%
  \BibitemOpen
 The definition we adopted is closer to algorithmic complexity. However, it is not concerned with the computer algorithm, and is with the diversity of chemical-reaction paths for polymerization.
  \bibitem [{\citenamefont {Trifonov}(1990)}]{trifonov1990making}%
  \BibitemOpen
  \bibfield  {author} {\bibinfo {author} {\bibfnamefont {E.~N.}\ \bibnamefont
  {Trifonov}},\ }\href@noop {} {\bibfield  {journal} {\bibinfo  {journal}
  {Structure and methods
  }\ } \textbf{\bibinfo {volume} {1}},\ \bibinfo
  {pages} {69-77} (\bibinfo {year} {1990})}\BibitemShut {NoStop}%
\bibitem [{\citenamefont {Popov}\ \emph {et~al.}(1996)\citenamefont {Popov},
  \citenamefont {Segal},\ and\ \citenamefont {Trifonov}}]{popov1996linguistic}%
  \BibitemOpen
  \bibfield  {author} {\bibinfo {author} {\bibfnamefont {O.}~\bibnamefont
  {Popov}}, \bibinfo {author} {\bibfnamefont {D.}~\bibnamefont {Segal}}, \ and\
  \bibinfo {author} {\bibfnamefont {E.~N.}\ \bibnamefont {Trifonov}},\
  }\href@noop {} {\bibfield  {journal} {\bibinfo  {journal} {BioSystems}\
  }\textbf {\bibinfo {volume} {38}},\ \bibinfo {pages} {65} (\bibinfo {year}
  {1996})}\BibitemShut {NoStop}%
\bibitem [{\citenamefont {Orlov}\ and\ \citenamefont
  {Potapov}(2004)}]{orlov2004complexity}%
  \BibitemOpen
  \bibfield  {author} {\bibinfo {author} {\bibfnamefont {Y.~L.}\ \bibnamefont
  {Orlov}}\ and\ \bibinfo {author} {\bibfnamefont {V.~N.}\ \bibnamefont
  {Potapov}},\ }\href@noop {} {\bibfield  {journal} {\bibinfo  {journal}
  {Nucleic Acids Res.}\ }\textbf {\bibinfo {volume} {32}},\ \bibinfo
  {pages} {W628} (\bibinfo {year} {2004})}\BibitemShut {NoStop}%
\bibitem [{\citenamefont {Shannon}(2001)}]{shannon2001mathematical}%
  \BibitemOpen
  \bibfield  {author} {\bibinfo {author} {\bibfnamefont {C.~E.}\ \bibnamefont
  {Shannon}},\ }\href@noop {} {\bibfield  {journal} {\bibinfo  {journal} {Bell Syst. Tech. J.}\ }\textbf {\bibinfo
  {volume} {27}},\ \bibinfo {pages} {379} (\bibinfo {year} {1948})}\BibitemShut
  {NoStop}%
\bibitem [{\citenamefont {Kolmogorov}(1963)}]{kolmogorov1963tables}%
  \BibitemOpen
  \bibfield  {author} {\bibinfo {author} {\bibfnamefont {A.~N.}\ \bibnamefont
  {Kolmogorov}},\ }\href@noop {} {\bibfield  {journal} {\bibinfo  {journal}
  {Sankhy{\=a}: The Indian Journal of Statistics, Series A}\ ,\ \bibinfo
  {pages} {369}} (\bibinfo {year} {1963})}\BibitemShut {NoStop}%
\bibitem [{\citenamefont {Lloyd}\ and\ \citenamefont
  {Pagels}(1988)}]{lloyd1988complexity}%
  \BibitemOpen
  \bibfield  {author} {\bibinfo {author} {\bibfnamefont {S.}~\bibnamefont
  {Lloyd}}\ and\ \bibinfo {author} {\bibfnamefont {H.}~\bibnamefont {Pagels}},\
  }\href@noop {} {\bibfield  {journal} {\bibinfo  {journal} {Ann.
  Phys.}\ }\textbf {\bibinfo {volume} {188}},\ \bibinfo {pages} {186}
  (\bibinfo {year} {1988})}\BibitemShut {NoStop}%
\bibitem [{\citenamefont {Gabrielian}\ and\ \citenamefont
  {Bolshoy}(1999)}]{gabrielian1999sequence}%
  \BibitemOpen
  \bibfield  {author} {\bibinfo {author} {\bibfnamefont {A.}~\bibnamefont
  {Gabrielian}}\ and\ \bibinfo {author} {\bibfnamefont {A.}~\bibnamefont
  {Bolshoy}},\ }\href@noop {} {\bibfield  {journal} {\bibinfo  {journal}
  {Comput. Chem.}\ }\textbf {\bibinfo {volume} {23}},\ \bibinfo
  {pages} {263} (\bibinfo {year} {1999})}\BibitemShut {NoStop}%
\end{thebibliography}

\begin{thebibliography}{2}%
\makeatletter
\providecommand \@ifxundefined [1]{%
 \@ifx{#1\undefined}
}%
\providecommand \@ifnum [1]{%
 \ifnum #1\expandafter \@firstoftwo
 \else \expandafter \@secondoftwo
 \fi
}%
\providecommand \@ifx [1]{%
 \ifx #1\expandafter \@firstoftwo
 \else \expandafter \@secondoftwo
 \fi
}%
\providecommand \natexlab [1]{#1}%
\providecommand \enquote  [1]{``#1''}%
\providecommand \bibnamefont  [1]{#1}%
\providecommand \bibfnamefont [1]{#1}%
\providecommand \citenamefont [1]{#1}%
\providecommand \href@noop [0]{\@secondoftwo}%
\providecommand \href [0]{\begingroup \@sanitize@url \@href}%
\providecommand \@href[1]{\@@startlink{#1}\@@href}%
\providecommand \@@href[1]{\endgroup#1\@@endlink}%
\providecommand \@sanitize@url [0]{\catcode `\\12\catcode `\$12\catcode
  `\&12\catcode `\#12\catcode `\^12\catcode `\_12\catcode `\%12\relax}%
\providecommand \@@startlink[1]{}%
\providecommand \@@endlink[0]{}%
\providecommand \url  [0]{\begingroup\@sanitize@url \@url }%
\providecommand \@url [1]{\endgroup\@href {#1}{\urlprefix }}%
\providecommand \urlprefix  [0]{URL }%
\providecommand \Eprint [0]{\href }%
\providecommand \doibase [0]{http://dx.doi.org/}%
\providecommand \selectlanguage [0]{\@gobble}%
\providecommand \bibinfo  [0]{\@secondoftwo}%
\providecommand \bibfield  [0]{\@secondoftwo}%
\providecommand \translation [1]{[#1]}%
\providecommand \BibitemOpen [0]{}%
\providecommand \bibitemStop [0]{}%
\providecommand \bibitemNoStop [0]{.\EOS\space}%
\providecommand \EOS [0]{\spacefactor3000\relax}%
\providecommand \BibitemShut  [1]{\csname bibitem#1\endcsname}%
\let\auto@bib@innerbib\@empty
\bibitem [{\citenamefont {Gardiner}(1985)}]{gardiner1985stochastic_sm}%
  \BibitemOpen
  \bibfield  {author} {\bibinfo {author} {\bibfnamefont {C.}~\bibnamefont
  {Gardiner}},\ }\href@noop {} {\emph {\bibinfo {title} {Stochastic Methods}}}\
  (\bibinfo  {publisher} {Springer-Verlag, Berlin--Heidelberg--New
  York--Tokyo},\ \bibinfo {year} {1985})\BibitemShut {NoStop}%
\bibitem [{\citenamefont {Van~Kampen}(1992)}]{van1992stochastic_sm}%
  \BibitemOpen
  \bibfield  {author} {\bibinfo {author} {\bibfnamefont {N.~G.}\ \bibnamefont
  {Van~Kampen}},\ }\href@noop {} {\emph {\bibinfo {title} {Stochastic Processes
  in Physics and Chemistry}}},\ Vol.~\bibinfo {volume} {1}\ (\bibinfo
  {publisher} {Elsevier},\ \bibinfo {year} {1992})\BibitemShut {NoStop}%
  \bibitem [{\citenamefont {Tkachenko}\ and\ \citenamefont
  {Maslov}(2017)}]{tkachenko2017onset_sm}%
  \BibitemOpen
  \bibfield  {author} {\bibinfo {author} {\bibfnamefont {A.~V.}\ \bibnamefont
  {Tkachenko}}\ and\ \bibinfo {author} {\bibfnamefont {S.}~\bibnamefont
  {Maslov}},\ }\href@noop {} {\bibfield  {journal} {\bibinfo  {journal} {arXiv
  preprint arXiv:1710.06385}\ } (\bibinfo {year} {2017})}\BibitemShut {NoStop}%
  \bibitem [{\citenamefont {Eigen}(1971)}]{eigen1971selforganization_sm}%
  \BibitemOpen
  \bibfield  {author} {\bibinfo {author} {\bibfnamefont {M.}~\bibnamefont
  {Eigen}},\ }\href@noop {} {\bibfield  {journal} {\bibinfo  {journal}
  {Naturwissenschaften}\ }\textbf {\bibinfo {volume} {58}},\ \bibinfo {pages}
  {465} (\bibinfo {year} {1971})}\BibitemShut {NoStop}%
\end{thebibliography}
\end{document}